\documentclass[amsmath,amssymb,prl,aps,showpacs,superscriptaddress,twocolumn]{revtex4-1}
\pdfoutput=1
\usepackage{amsmath,amsfonts,amssymb,amsthm,graphics,graphicx,epsfig,bbm}
\usepackage[colorlinks=true,citecolor=blue,linkcolor=blue,urlcolor=blue]{hyperref}
\usepackage[usenames]{color}
\usepackage{graphicx}
\usepackage{subfigure}
\usepackage{amsmath}
\usepackage{epsfig}
\usepackage{dcolumn}
\usepackage{bm}
\usepackage{color}
\usepackage{epstopdf}
\usepackage{amssymb}
\usepackage{amstext}
\usepackage{latexsym}
\usepackage{hyperref}
\usepackage{amsfonts}
\usepackage{psfrag}
\usepackage{xcolor}
\usepackage[normalem]{ulem}
\usepackage{dsfont}
\usepackage{txfonts}
\usepackage{cases}
\usepackage{ifthen}
\usepackage{algorithm}
\usepackage{algpseudocode}

\graphicspath{{./Images/}}

\newcommand{\ket}[1]{\ensuremath{\left\vert #1 \right\rangle}}

\begin{document}
\title{Multi-mode N00N states in driven atomtronic circuits}

\author{Enrico Compagno}
\affiliation{Univ. Grenoble-Alpes, CNRS, Institut N\'eel, 38000 Grenoble, France}

\author{Guillaume Quesnel}
\affiliation{Univ. Grenoble-Alpes, CNRS, Institut N\'eel, 38000 Grenoble, France}

\author{Anna Minguzzi}
\affiliation{Univ. Grenoble-Alpes, CNRS, LPMMC, 38000 Grenoble, France}

\author{Luigi Amico}
\affiliation{Dipartimento di Fisica e Astronomia 'Ettore Majorana', Via S. Sofia 64, 95127 Catania, Italy}
\affiliation{CNR-IMM $\&$ INFN-Sezione di Catania, Via S. Sofia 64, 95127 Catania, Italy}
\affiliation{Centre for Quantum Technologies, National University of Singapore,
3 Science Drive 2, Singapore 117543, Singapore}
\affiliation{LANEF Chaire d'excellence, Univ.  Grenoble-Alpes \& CNRS, F-38000 Grenoble, France}

\author{Denis Feinberg*}
\affiliation{Univ. Grenoble-Alpes, CNRS, Institut N\'eel, 38000 Grenoble, France}

\date{\today}


\begin{abstract}
  We propose a method  to generate multi-mode N00N states with arrays of ultracold atoms.
  Our protocol requires a strong relative offset among the wells
  and a drive of the interparticle interaction at a frequency resonant with the offsets. The proposal is
  demonstrated by a numerical and a Floquet analysis of the quantum dynamics of a ring-shaped atomtronics circuit made of M weakly coupled optical traps. We generate a hierarchy of energy scales down to very few low-energy states where N00N dynamics takes place, making multi-mode N00N states appear at nearly
  regular time intervals.
  The production of multi-mode N00N states can be probed by time-of-flight imaging.  Such states may be used to build a multiple beam splitter.
\end{abstract}
\maketitle

\section{I. Introduction.}
Entanglement is at the root of quantum technology \cite{dowling2003quantum}. 
Depending on the number of subsystems involved, such genuine quantum correlations can be of very different nature \cite{horodecki2009quantum,amico2008entanglement}. While the picture for the  pairwise case has been cleared up, multipartite entanglement remains challenging for both its mathematical analysis and quantum state engineering (see \cite{dur_vidal_cirac2000,johansson2014classification,neeley2010generation}).  Nonetheless, multipartite entangled states were recognised  as an important  resource in different areas of quantum technology, including quantum information theory, quantum cryptography and noteworthy quantum sensing \cite{walter2016multipartite,pezze2018rmp,caves_prd1981,treutlein_science2018}.
As an example, non-classical states of several identical particles
possess intrinsic quantum correlations that can be exploited in technological applications. Among the most well-known non-classical states,  
the bipartite ``N00N'' states correspond to the macroscopic superposition $\psi=(|N,0\rangle+ |0,N\rangle)/\sqrt{2}$ where $|n_1,n_2\rangle$ have $n_1$, $n_2$ bosons in the modes $1$ and $2$ respectively. Such states are maximally entangled and highly sought for applications e.g. to high-precision interferometry \cite{dowling2008,ono2013,jones2009}. 

We focus on N00N states with multipartite entanglement \cite{humphreys2013}. The suggested platform is provided by cold atoms. 
They can be controlled with high precision and flexibility of the operating conditions \cite{acin2018quantum}. In these systems, the physical conditions, as for instance the atom-atom interaction, can be tuned in a wide range of values or as function of time \cite{gong_many-body_2009,rapp_ultracold_2012,watanabe_floquet_2012,Creffield2014,meinert_floquet_2016}. 
Control of the atom number in small traps is achieved by fluorescence \cite{oberthaler2015}. It is now possible to handle the atomic confinement on local spatial scales (micrometers)  \cite{henderson_experimental_2009,rubinsztein2016roadmap}, and on time scales much shorter than atom's dynamics time scales \cite{lignier2007dynamical,alberti2009engineering,zenesini2009coherent,gauthier_direct_2016,jiang_majorana_2011,
Liang:09,muldoon_control_2012,henderson_experimental_2009}.
Atomtronics exploits these progresses with the goal to widen the scope of cold atom simulators and to achieve new concepts of devices of practical value \cite{dumke_roadmap_2016,amico_focus_2017}.
Roads to formation of bipartite N00N states with ultracold atoms were proposed either by phase imprinting followed by a free evolution \cite{cirac1998,mahmud2005,leung2012} , or via rotation of the condensate \cite{hallwood2006}, or using time-dependent fields \cite{stiebler2011,zhou2013,watanabe_floquet_2012,yukawa2018}. 
Very few experiments with N00N states exist with ultracold atoms \cite{chen2010},  and other non-classical states have been reported such as squeezed \cite{pezze2018rmp} and over-squeezed states \cite{nascimbene2019}.

\begin{figure}[htbp]
\centering
\includegraphics[width=0.75\columnwidth]{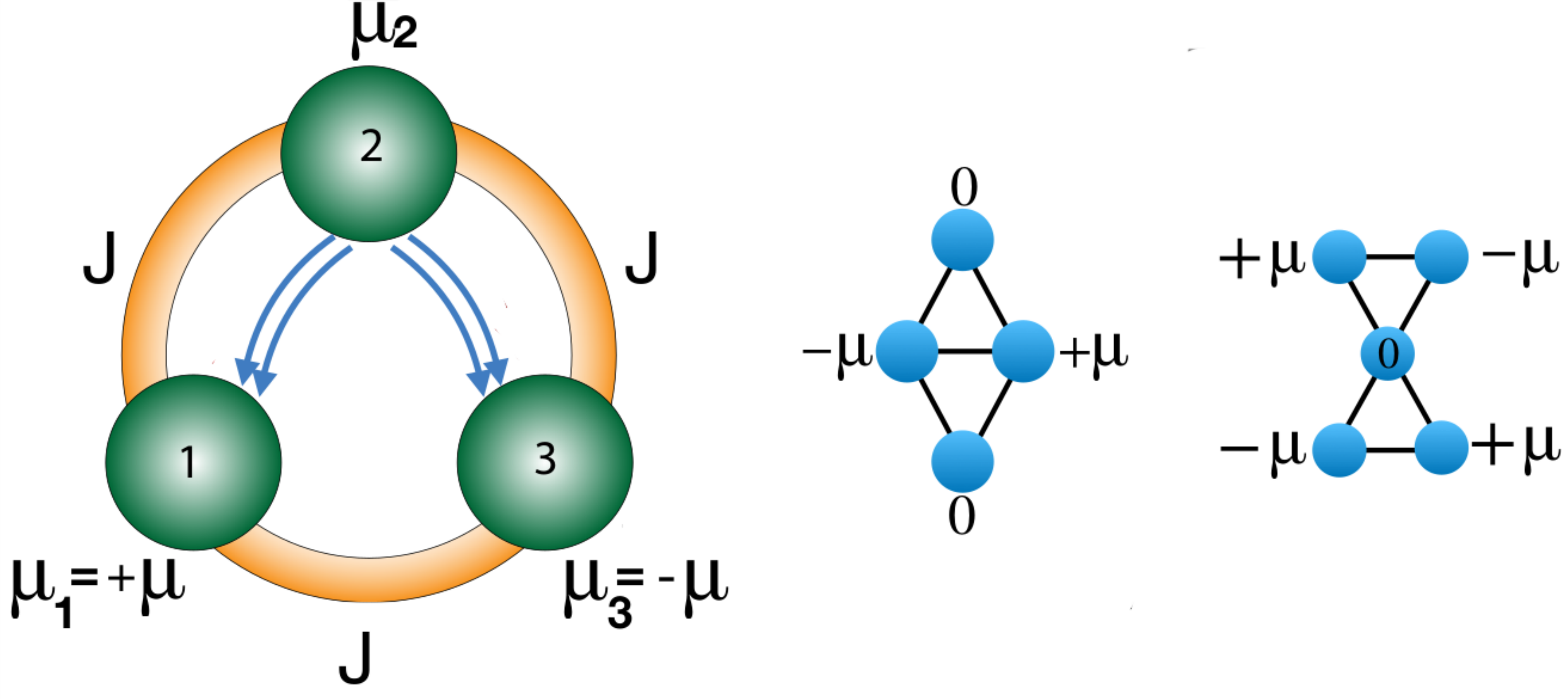}
\caption{\small{(Color online) Main figure : Scheme of the three-well geometry:
    bosons are confined  in a triple-well ring configuration, where $J$ is the tunnelling rate, $\mu_1=\mu$, $\mu_2=0$ and $\mu_3=-\mu$ are respectively the local energy
    bias
    in sites $1,2$ and $3$. The blue arrows represent  motion of bosons in pairs, triggered by the periodic drive. Insets: $M=4$-sites and $M=5$-sites configurations with corresponding bias.}}
\label{fig:Scheme3BECRing}
\end{figure}

In this work, we propose a protocol for the generation of multi-mode N00N states, i.e. the states of the type $|N000...\rangle+ |0N00...\rangle+ |00N...\rangle+...$. Such states are relevant to different contexts of quantum technology such as multiport interferometry \cite{humphreys2013}. Here we show 
how the interplay between periodic driving of the interaction and a strong offset can generate multimode N00N states with high fidelity at nearly periodic time intervals. As examples, we consider mesoscopic atom numbers (from $N=2$ to $10$) 
and number of modes $M=2-5$.

Our atomtronic circuit is made of $N$ neutral bosons trapped in $M$ optical traps
arranged in a ring geometry at zero temperature (Fig.\ref{fig:Scheme3BECRing}).
All the well's minima are offset by an energy $\pm \mu$
with respect to 
a reference well (the ``source''), with $\mu\gg J_{ij}$ where the $J_{ij}$ are the inter-well tunneling amplitudes. The interaction is driven periodically \cite{strzys2008,
  gong_many-body_2009,watanabe2010,rapp_ultracold_2012,watanabe_floquet_2012,Creffield2014,meinert_floquet_2016} around $0$ with frequency $\omega$, chosen to be resonant with the offset, $\hbar \omega=2\mu$. This makes both single-atom transitions between offset wells and pair transitions from the source well resonant, with very different energy scales. A Floquet analysis shows that this generates very few low-energy states connecting the N00N state components. Optimal parameters for N00N states are found close to Coherent Destruction of Tunnelling (CDT \cite{grossmann1991coherent,grifoni_driven_1998,gong_many-body_2009,watanabe_floquet_2012}) from the initial state, which we explain by a "Russian doll" fitting of five different energy (or time) scales, down to the N00N recurrence time. This turns out to be a powerful scenario to create multi-mode N00N states ($M\geq3$) on time scales much shorter than those associated to the free evolution of the system (e.g. after a quench) \cite{zoller2003,ferrini2008,smerzi2008}. Our analysis involves an exact solution of the quantum dynamics from the time-dependent Hamiltonian, as well as a high-frequency expansion of the Floquet Hamiltonian, that gives excellent agreement with the direct solution. 
Finally, we discuss how to detect experimentally such a correlated transport: second-order  momentum correlations, as obtained after a time-of-flight expansion, can distinguish between coherent and  localized states achieved in the transfer. 

Section II presents the model and the analytical or numerical methods. Section III presents in detail the correlated boson transfer obtained from the driven dynamics. Section IV discusses the setup constraints and provides an explanation of multipartite N00N state formation. Section V explains the readout protocol. Section VI concludes, and a few Appendices provide more details. 

\section{II. The model and its analysis.}
The system is described by the M-mode Bose-Hubbard Hamiltonian subjected to a periodic modulation of the interaction strength:
\begin{eqnarray}
\nonumber
H(t)&=&-\sum\limits_{i,j=1(i\neq j)}^{M}\left(J_{ij}\,a_{j}^\dagger\,a_i  + H.c.\right) + \\
&+&\big(U_0+\frac{U_1}{2}\sin(\omega t)\big) \sum\limits_{i=1}^{M} \hat{n}_i\,(\hat{n}_i-1) - \sum\limits_{i=1}^{M}\mu_i \,\hat{n}_i ~,
\label{eq:BoseHubbardRing}
\end{eqnarray}
where $a_i^\dag$ creates a boson in site $i$ and $\hat{n}_i=a_i^\dag a_i$ is the number operator. The parameters $J_{ij}$ quantify the hopping strengths, $U_0$ is the average inter-particle interaction, 
$U_1$ the amplitude of its periodic modulation and $\mu_i$ the local chemical potential, i.e. the well offset of the site $i$. 

Let us first briefly mention the behavior of a three-atom ring with strong offset but static interaction $U_0$ only and $\mu_1=\mu,\mu_3=-\mu$. Details can be found in Appendix A. A second-order perturbative analysis in $J/\mu$ shows how to transfer a single pair with high fidelity from site $2$ to sites $(1,3)$ altogether, i.e. obtain a nearly perfect Rabi resonance between states $\ket{n_1,n_2,n_3}$ and $\ket{n_1+1,n_2-2,n_3+1}$. Such a highly correlated transition requires a fine tuning of the offset $\mu_2$. Importantly, this does not allow to transfer many pairs simultaneously, and the presence of a strong residual interaction that competes with the pair hopping prevents from creating a coherent state made of boson pairs.

Our aim is now to investigate 
how a correlated pair transport of ultra-cold bosonic atoms can be triggered with $U_0=0$ by a suitable choice of $\mu_i$, $U_1$ and $\omega$. 
To illustrate the analysis in presence of a driven interaction, let us consider a three-well ring (see Fig.~1), with equal and real couplings $J_{ij}=J$, and offsets $\mu_{1,2,3}=\mu, 0, -\mu$. We drive the interaction strength as in Eq. (\ref{eq:BoseHubbardRing}) with $\omega\gg J/\hbar$ (see \cite{bermudez2011synthetic,kolovsky2011creating}). 
A key parameter for the dynamics is $K_0=U_1/(\hbar \omega)$. Given the values of the relative offsets between the bosonic islands  $(1,2,3)$, 
one encounters two different Josephson frequencies for noninteracting atoms in a ring geometry: $\omega_J=\mu/\hbar$ for single boson transitions between $1$ (or $3$) and $2$, 
and $\omega_J=2\mu/\hbar$ for boson pairs hopping between $1$ (or $3$) and $2$, as well as for single boson transitions between $1$ and $3$. 
Therefore, coherent pair correlations are  expected to be formed for resonant driving  frequencies $\omega=2\mu/\hbar$. 

We study the dynamics through exact diagonalization of the many-body Hamiltonian (\ref{eq:BoseHubbardRing}) and the evolution of an initial state chosen to be $\ket{0N0}$.
We complement this study by the analysis of an effective time-averaged Hamiltonian, following the method proposed by 
Dalibard and Goldman \cite{goldman2014periodically,goldman2015periodically} (see also Ref. \onlinecite{eckardt_high-frequency_2015}). Our derivation (Appendix B) involves 
elimination of the second and third terms in Eq. (\ref{eq:BoseHubbardRing}) by a canonical transformation. The dressed $1\leftrightarrow3$ transitions can be directly 
averaged on time while the $1\leftrightarrow2$ and $3\leftrightarrow2$ transitions require a Floquet expansion of order ${1}/{\omega}$ yielding: 
\begin{widetext}
\begin{eqnarray}
\nonumber
H_{eff}&=&-J\,\Big[\mathcal{J}_1[K_0(\hat{n}_1-\hat{n}_3-1)]\,a_1^\dagger\,a_3\,+\,H.c.\Big]\,+\,\frac{J^2}{\mu}\,\Big[\mathcal{L}(K_0,\hat{n}_1,\hat{n}_2)\,(a_1^\dagger a_2)^2
+\mathcal{L}(K_0,\hat{n}_2,\hat{n}_3)\,(a_2^\dagger \,a_3)^2\, \\
&&+\,\mathcal{M}(K_0,\hat{n}_1,\hat{n}_2,\hat{n}_3)\,a_1^\dagger\, a_3^\dagger\, (a_2)^2\,
+\mathcal{N}(K_0,\hat{n}_1,\hat{n}_2,\hat{n}_3)\,a_1^\dagger\,a_3\,+\,H.c.\Big]\\
\nonumber
&&+\frac{J^2}{\mu}\,\Big[\hat{n}_2(\hat{n}_1+1)\,\mathcal{P}(K_0,\hat{n}_1,\hat{n}_2)-\hat{n}_1(\hat{n}_2+1)\,\mathcal{P}(K_0,\hat{n}_2,\hat{n}_1)
+\hat{n}_2(\hat{n}_3+1)\,\mathcal{P}(K_0,\hat{n}_2,\hat{n}_3)-\hat{n}_3(\hat{n}_2+1)\,\mathcal{P}(K_0,\hat{n}_3,\hat{n}_2)\Big]\\
\nonumber
\label{eq:effectivemodel}
\end{eqnarray}
\end{widetext}
$\mathcal{L},\mathcal{M},\mathcal{N},\mathcal{P}$ being infinite series of Bessel functions of the arguments $K_0(\hat{n}_i-\hat{n}_j)$  (see Appendix B).

\section{III. Correlated atom pair transport.}
The initial state involves $N$ atoms in the unbiased trap $2$. This choice aims at a redistribution in time of the N atoms into traps $1$ and $3$.  Traps $2$, and $1,3$ can be respectively considered as the "source" and "output" modes of an atomic "beam splitter". 
The exact evolution of the initial state $\Psi(t=0)=\ket{0N0}$ under the time-dependent Hamiltonian, or under the Floquet expansion Hamiltonian $H_{eff}$, is obtained by discretization of the Trotter formula. Convergence is checked with respect to the time step, much smaller than  the smallest time scale in the Hamiltonian i.e. the drive period. 

 \subsection{The two-mode model.}
\begin{figure}[htbp]
\centering
\includegraphics[width=0.8\columnwidth]{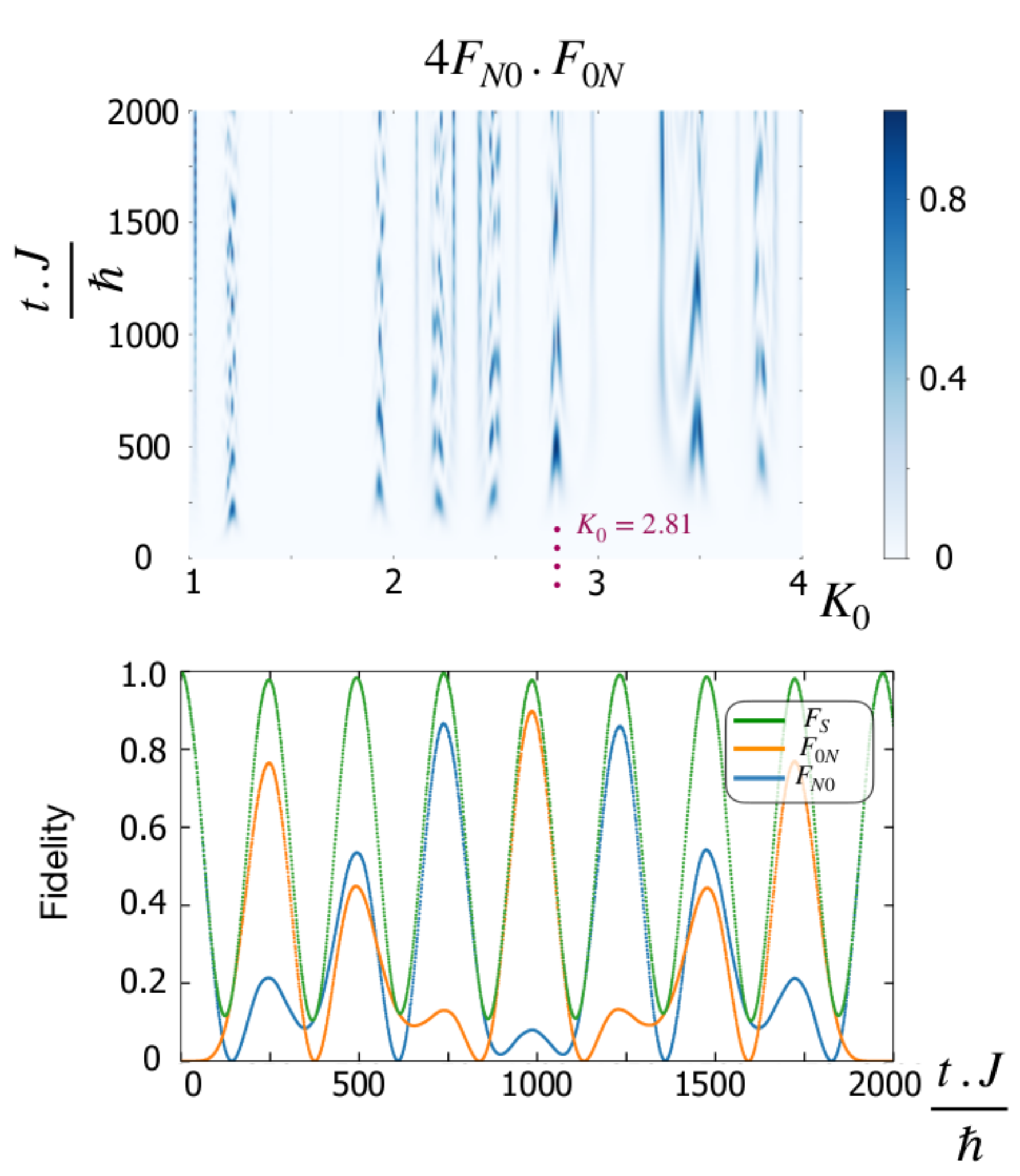}
\caption{\small{(Color online)
Generation of two-mode N00N states. (Top) Map of the N00N fidelity indicator $4F_{N0}F_{0N}$ with the interaction $K_0$ and reduced time;
 (Bottom) Time variation of the fidelities for a N00N state ($F_S=F_{N0}+F_{0N}$). $N=10$ particles, $\mu=15 J$, $K_0=2.81$.}}
\label{fig:N00N}
\end{figure}

Let us benchmark our protocol with a driven two-mode model. In contrast with the models studied so far \cite{mahmud2005,leung2012,watanabe_floquet_2012}, driving of the interaction is combined with a large static offset. Starting in the state $\ket{N,0}$ at time $t=0$, Fig. \ref{fig:N00N} show a fidelity map for the N00N state as a function of $K_0$ and reduced time $t$ (in units of $\hbar/J$). We choose as an indicator the product $4F_{N0}F_{0N}$ where $F_{N0}=|\langle N0|\Psi(t)\rangle|^2$,  $F_{0N}=|\langle0N|\Psi(t)\rangle|^2$. It shows that N00N states are created in sizeable intervals of $K_0$ (Fig.~\ref{fig:N00N}), nearly with time periodicity. The recurrence time of $N00N$ states is controlled by the interplay of small energy scales and is of the order of a few hundred times $\hbar/J$. The protocol operates for even values of $N$.
Its efficiency comes from the atom pair dynamics triggered by the resonant drive, which considerably reduces the effective Hilbert space and helps atoms to bunch equally in both wells. 

\subsection{Three modes and tripartite entanglement.}
Remarkably, the two-mode result generalizes to larger well arrays, despite the growth of the Hilbert space. Let us consider the three-well symmetric case. 
Eq. (2) shows that atom pair transfers  $2 \leftrightarrow 1$  and $2 \leftrightarrow 3$ as well as $2 \leftrightarrow (1,3)$ are dynamically generated.

\begin{figure}[t]
\includegraphics[width=0.8\columnwidth]{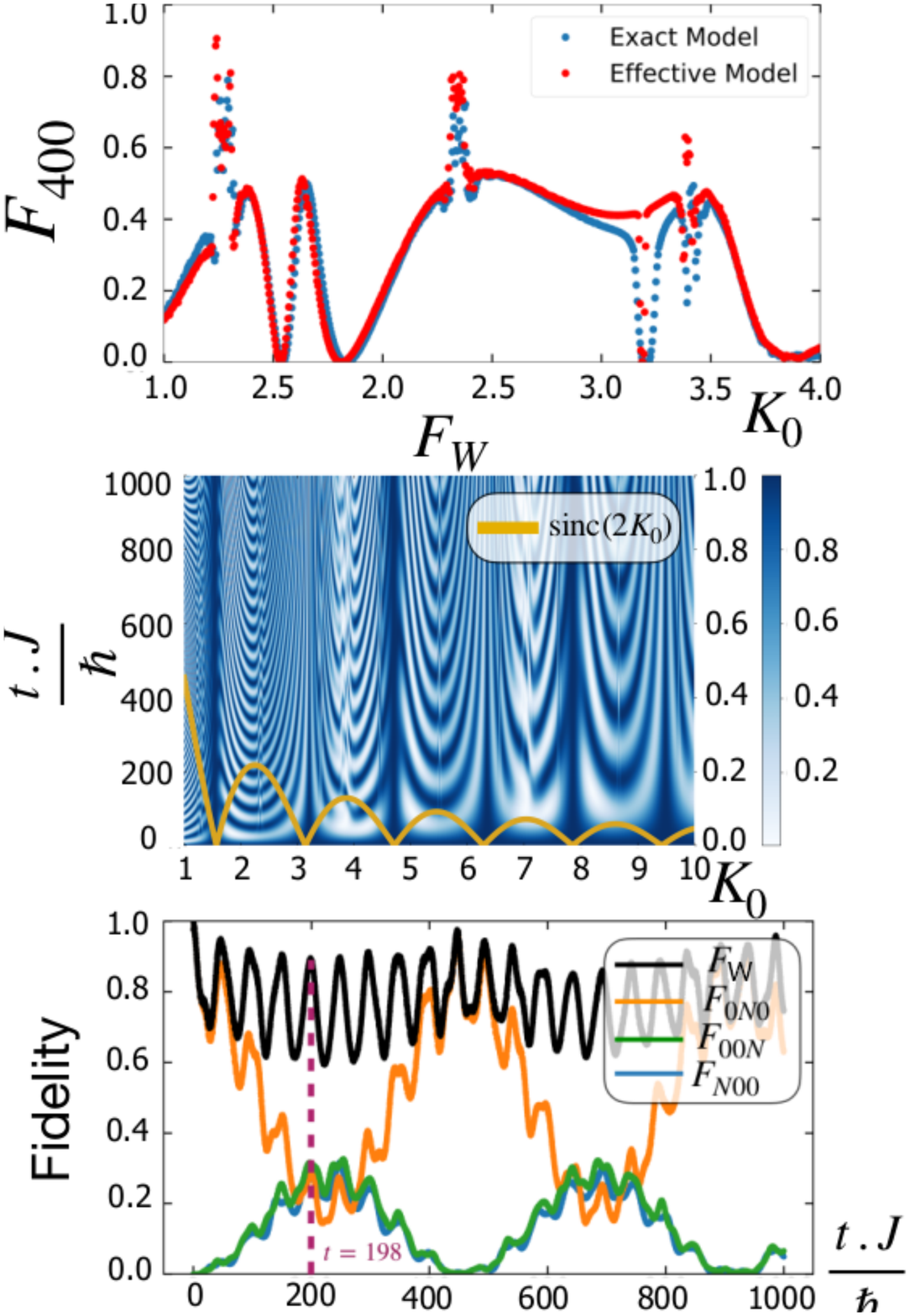}
\caption{\small{(Color online) Optimum transfer and three-mode N00N states. (Top) Comparison of the fidelity (maximized over $t\sim 2000\,\hbar /J$) for transfer from $\ket{0,4,0}$ to  $\ket{4,0,0}$, calculated from the exact dynamics (blue) and from the Floquet expansion (red). (Middle) Interaction-time map of the fidelity $F_W$. (Bottom) Time evolution of the fidelities $F_W$ (in black) and for states $\ket{N,0,0}, \ket{0,N,0} , \ket{0,0,N}$, with $N=8$ and $K_0=2.59, \,\mu=17J$.}}
\label{fig:Exact_vs_Eff}
\end{figure}

{\it Directed transfer of N atoms.} Let us first consider full transfer from the ``source'' trap $2$ to trap $1$. 
Fig. \ref{fig:Exact_vs_Eff} shows the time dynamics of the transfer fidelity $F_{N00}=|\langle N00|\Psi(t)\rangle|^2$ from state $\ket{0,N,0}$ to state $\ket{N,0,0}$ as a function of the interaction, as well as its maximum on a large time interval (middle and top panels).
 The agreement  between the exact dynamics and the one resulting from the effective Hamiltonian Eq.~(\ref{eq:effectivemodel}) is remarkable.
For specific values of $K_0$, the fidelity displays marked peaks.  
Such a targeted optimal transfer requires a fine tuning of the interaction parameter, especially for large (even) $N$. The system achieves a coherent transfer of $N$ bosons from trap $2$  to trap $1$ (or to trap  $3$, at the same $K_0$ value but at different times). Moreover, for  $K_0 \gg 1$,  the asymptotic expansion of the Bessel functions gives $\mathcal{L} \sim \frac{\sin(2K_{0})}{2K_0}$ (Appendix B).  Therefore the transfer of pairs from the islands  $2$ to $1$ or $3$ is suppressed for   $K_{0,n} \sim {n\pi}/{2} ,\;  n\in\mathbb{N}$. One verifies that this makes the system trapped in the initial state $\ket{0 , N , 0}$, by a CDT mechanism \cite{grossmann1991coherent}, as visible on the fidelity map of Fig. \ref{fig:Exact_vs_Eff}. 

{\it Three-mode generalized N00N states.} We now demonstrate that, exploiting the ring geometry,  coherent superpositions of the states $\ket{N,0,0}, \ket{0,N,0}, \ket{0,0,N}$ can be dynamically created for a wider range of system's  parameters. 
To illustrate this, we represent the state of the system as $|\Psi_S(t)\rangle \,= |\Psi_W (t)\rangle + |\Psi_{W \perp} (t) \rangle $ with:
\begin{equation}
\label{eq:Wstate}
|\Psi_W(t)\rangle =\,a_1(t)\ket{N,0,0}\,+\,a_2(t)\ket{0,N,0}\,+\,a_3(t)\ket{0,0,N} 
\end{equation}
with $|\Psi_{W\perp} (t)\rangle$ an orthogonal vector to $|\Psi_{W}(t)\rangle$, 
and compute  the fidelity on
$|\Psi_S(t)\rangle$
\begin{equation}
F_W(t)\,=\,|a_1(t)|^2\,+\,|a_2(t)|^2\,+\,|a_3(t)|^2
\label{eq:fidelity}
\end{equation}
For all the explored values of $N$ ($2\leq N\leq 10$),  we find that $F_W(t)$ maxima are nearly periodic and close to $1$ (see Fig. \ref{fig:Exact_vs_Eff}, bottom panel and Appendix C).  In particular,  in specific  intervals of  $K_{0}$,  a three-component $W$-like superposition $\Psi_{W}$ \cite{dur_vidal_cirac2000} with comparable amplitudes $|a_1|\simeq|a_2|\simeq|a_3|$ is generated  with high fidelity at nearly regular times. Notice that forming a W-state $a_1\ket{0N0} + a_2\ket{N00}+a_3\ket{00N}$ is equivalent to a 3-mode beam splitter operation. 

One can check the effect of asymmetries (see Appendix C). First, taking the couplings $J_{21}$ and $J_{23}$ from the ``source'' well $2$ as different, a similar efficiency is achieved for moderate asymmetry $\frac{|J_{21}-J_{23}|}{J_{21}+J_{23}}$. Second, varying the coupling $J_{13}$ has a stronger effect. If $J_{13}=0$, i.e. a linear configuration of the three wells, no three-mode N00N state is achieved, only two-mode N00N states are attained, either between wells $2,1$ or between wells $2,3$. Strikingly, the ring configuration turns out to be essential to create multimode N00N states.

\subsection {More than three modes.}
\begin{figure}[htbp]
\centering
\includegraphics[width=0.8\columnwidth]{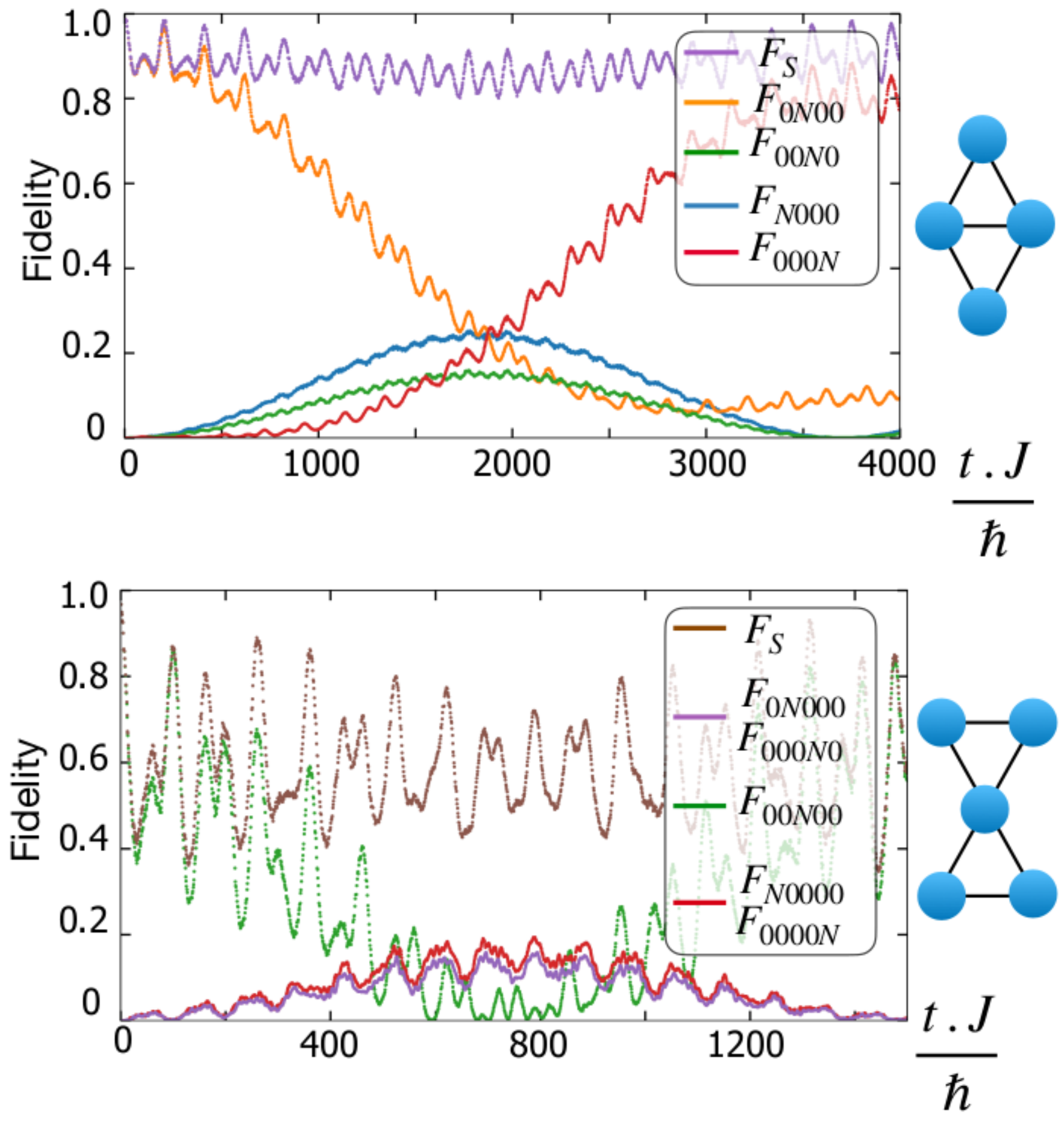}
\caption{\small{(Color online)
Generation of M-mode N00N states. (Top) Four sites, $K_0=3.08$, $\mu=20$, $N=4$; (Bottom) Five sites, $K_0=1.57$, $\mu=15$, $N=4$. The total fidelity $F_S$ generalizes Eq. \ref{eq:fidelity} to $M$ components.}}
\label{fig:M-N00N}
\end{figure}

Remarkably, our protocol can be generalized to engineer $M$-mode W-states with $M>3$. A four-mode and a five-mode $N00N$ generator circuit are represented on Fig. 1. The latter is scalable to any odd $M$. Fig. \ref{fig:M-N00N} shows fidelities up to $90\%$ for $M=4$ and $80\%$ for $M=5$.
For $N$ atoms in $M$ modes the size of the Hilbert space is
$
{\cal N}(N,M)\,=\,\frac{(N+M-1)!}{N!(M-1)!}
$
e.g. for instance ${\cal N}(4,3)=15$, ${\cal N}(10,3)=66$, ${\cal N}(4,5)=70$. Achieving N00N states with the latter numbers requires a precision of about $1\%$ in the choice of parameter $K_0$. This precision grows with ${\cal N}(N,M)$ but the average time (recurrence time) $\Delta t$ separating the occurrence of N00N states does not increase. For the values of $\mu$ taken in this study, the recurrence time is of the order of a few hundreds times $\frac{\hbar}{J}$. This time scales linearly with the ratio $\frac{\mu}{J^2}$, as shown by the amplitude of the pair atom transfer in the Floquet expansion (Eq. \ref{eq:effectivemodel}).

 \section{IV. Discussion.}
\subsection{Energy scale separation.}
The formation of multimode N00N states stems from the Floquet effective model, Eq. (2). It relies on energy scale separation. The resonant drive eliminates the largest scale $\mu$ from the averaged motion, leaving a dressed hopping of order $J$ between wells $1,3$, and pair hoppings as well as potential terms of order $J^2/\mu<<J$. As a result, the spectrum of $H_{eff}$ shows a few low-energy states with splittings $\sim J^2/\mu$, among high-energy states at scale $\sim J$. 
During time evolution from state $\ket{0N0}$ ($N$ even) the wavefunction mostly stays within the low-energy sector {\it made of paired states} $\ket{n_1,n_2,n_3}$ i.e. with $n_i$ even, while most Fock states other than $\ket{N00}$, $\ket{00N}$ participate in the high-energy one. This eventually makes the fully unbalanced states $\ket{N00..}$, $\ket{0N0..}$ linked together by a connecting chain of about $\frac{N}{2}$ states {\it only}. This drastically reduces the transition time between those states, compared to an undriven system. The process can be further optimized by noting that close to values yielding CDT, $K_{0}\simeq n\frac{\pi}{2}+\delta K_0$, some of the transition amplitudes within the connecting chain are very small, of order $\delta J = (J^2/\mu)\delta K_0/K_0$. A perturbative analysis eventually makes a much lower energy scale emerge, that controls the transition between states $\ket{N00..}$, $\ket{0N0..}$,...  To summarize, a kind of renormalization scheme produces a hierarchy of energy scales $\mu \gg J \gg J^2/\mu \gg \delta J \gg (\delta J/J)^2 \mu \sim \hbar/\tau_{rec}$, where $\tau_{rec}$ is the typical N00N recurrence time (see Appendix D for a detailed analysis). The latter time is still experimentally accessible, as it is way shorter than the recurrence time predicted from a standard undriven dynamics of an interacting system. 

\subsection{The setup constraints.}
Let us now comment on the setup constraints. First, the multi-mode N00N states generated at times $t_i$ are perfectly coherent. Yet, the relative phases of the components of each state change from a time $t_i$ to another recurrence (see Appendix C). Notice that this is not detrimental to the multipartite entanglement of each state, and allows its use as a multi-phase probe after release. Second, our protocol requires even filling of the trap array, and the atom number to be constant during a time interval $\Delta t$, where $\Delta t \simeq|t_{i+1}-t_i|$. This requires an evaporation time $\tau_{ev}>\Delta t$. 

The protocol assumes a drive resonant with the offsets. If this resonance is not perfect, i.e. $\omega=2\mu/\hbar+\delta \omega$, the many-body interference making multi-mode N00N states is destroyed at long times. This can be estimated by comparing the "beating" frequency $\delta \omega$ to the recurrence time $\Delta t$. We indeed find numerically that N00N states still occur at times $t << \delta \omega^{-1}$, which requires the condition $\delta \omega << \Delta t^{-1}$ for the protocol to operate. This implies in practice that the offset $\mu$ should not be taken too large.

\section{V. Readout of multi-mode N00N states.} 
\begin{figure}[t]
\includegraphics[width=0.8\columnwidth]{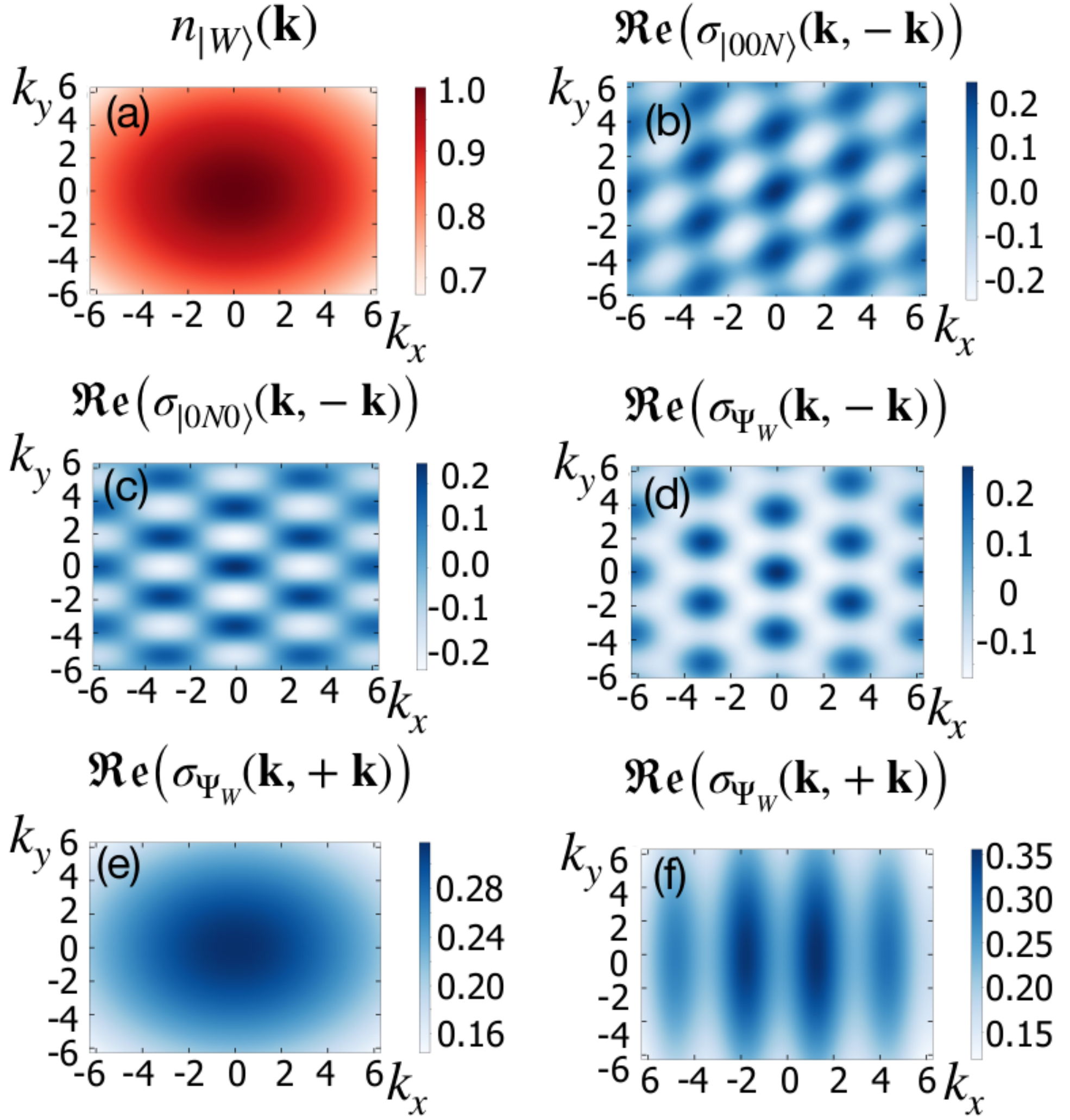}
\caption{\small{(Color online) Read-out maps (a) $n({\bf k})$ for a $W$-state; (b) $\sigma({\bf k},-{\bf k})$ for $\ket{0,0,N}$, (c) $\sigma({\bf k},-{\bf k})$ for $\ket{0,N,0}$, (d) $\sigma({\bf k},-{\bf k})$ for the (nonideal) $W$-type state; $\sigma({\bf k},{\bf k})$ for a $W$-type state, either perfect (e) or with a component $|\Psi_{W \perp} (t) \rangle$ (Eq. \ref{eq:Wstate})  (f). $N=8$.}}
\label{fig:readout}
\end{figure}

The diagnostic of the system relies on the time-of-flight technique:  in our dynamical scheme, the state 
is prepared,  driven and then released at the time $t=\bar{t}$. The long-time density  in real space can be accessed by the Fourier transform at the releasing time $\bar{t}$  \cite{bloch2008zwerger,amico_quantum_2005} 
\begin{equation}
\hat n({\bf k})\, = \,\sum_{i,j}\,e^{i\bf k\cdot ({\bf R}_i-{\bf R}_j)} a^{\dagger}_ia_j ,\;\;\;{n}({\bf k},\bar{t})=\,\langle \Psi(\bar{t}) |\,\hat n({\bf k}) \,| \Psi(\bar{t}) \rangle 
\end{equation}
While $n({\bf k})$ is featureless (Fig. \ref{fig:readout}a), pair and many-boson pair transfers can be probed through the correlations in $\hat n({\bf k})$:  
\begin{equation}
\sigma({\bf k},{\bf k'},\bar{t})\, = \,\langle \Psi_S(\bar{t}) |\,\hat n({\bf k})\,\hat n({\bf k'})\,| \Psi_S(\bar{t}) \rangle
-\,{n}({\bf k},\bar{t})\,{n}({\bf k'},\bar{t})
\end{equation}
A perfect $\ket{0N0}$ state (Fig.~\ref{fig:readout}c) is characterized by a pattern in $\sigma({\bf k,\bf k'})$ with $\bf k=-\bf k'$, due to coherent virtual single-atom transitions towards one of the unoccupied sites  (Appendix E). 
Rotated but similar patterns are obtained for states $\ket{0,0,N}$ (Fig.~\ref{fig:readout}b) and $\ket{N,0,0}$. The combination thereof  gives rise to the pattern corresponding to the optimal W state:  $\frac{1}{\sqrt{3}}(\ket{0,N,0}+\ket{N,0,0}+\ket{0,0,N})$. The pattern for a dynamically achieved $W$-type state (see Fig. \ref{fig:readout}d) is indeed very similar to that of an optimal one (see Appendix E).  
Remarkably, it is also possible to probe the overall fidelity of the $W$-state: indeed, the patterns taken instead at $\bf k=\bf k'$ are structureless for ideal $W$- states. Therefore they allow probing other states contributing to $F_W<1$ (Fig.~\ref{fig:readout}e-f) (see Appendix E for details). 

\section{VI. Conclusions.} 
We have considered a class of atomtronic circuits to study the correlated dynamics in bosonic networks: M coupled mesoscopic dots of interacting particles in -- this is essential -- a ring geometry.  The correlated transfer results from the combination of suitable offset potentials and strong resonant driving of the interaction.
The dynamics of such a system is dictated by short chains of states connecting the N00N components, making multi-mode N00N states form on realistic time scales. The diagnosis of the system states is carried out through the analysis of the momentum distribution which, in a cold atoms setting, corresponds to  time of  flight images. Our protocol continously produces multi-mode N00N states and is only limited by the lifetime of the optically trapped atoms.  
These states could be used to bring parallelism in multiple phase-imaging protocols \cite{humphreys2013}. 
Our study  provides a new route to engineer  complex correlations in quantum networks. The physical system we studied is within the current experimental capabilities in the atomtronics field.

{\it Acknowledgments}
E. C. thanks S. Bose and L. Banchi for interesting
discussions and suggestions. A. M. and D. F. thank H. Perrin for useful insights on the experimental feasibility of the setup. We acknowledge the ANR SuperRing project ANR-15-CE30-0012-02 and the Grenoble LANEF framework ANR-10-LABX-51-01 for their support with mutualized infrastructure. 

*Corresponding author: denis.feinberg@neel.cnrs.fr

\section{Appendix A: Nonlocal pair transport combining bias and static repulsion.}
Let us consider the effect of the strong offset $\mu$ with a {\it static} interaction only, with the Hamiltonian:
\begin{equation}
H_=\frac{U_0 }{2} \sum\limits_{j=1}^{3} n_j(n_j-1) - \sum\limits_{j=1}^{3}\mu_j n_j\,
-\,J\sum\limits_{i,j=1(i\neq j)}^{3} \left(a_j^\dag a_i + H.c.\right)~.
\label{eq:BoseHubbardPerturbative}
\end{equation}
This simple system indeed triggers correlated pair hopping between site $2$ and sites ($1,3$) altogether. 
We choose for this purpose $\mu_1=-\mu_3=\mu$. Specifically, starting from the initial state $\ket{\psi_{i}}=\ket{n_{1i},n_{2i},n_{3i}}$, the correlated transport of $m$ pairs brings the system coherently through the states $\ket{\psi_f}=\ket{n_{1i}+{m},n_{2i}-2{m},n_{3i}+m}$. $N_T$ being the total number of pairs, an entangled target state is a superposition of such states:
\begin{equation}
	\ket{\psi_T} = \sum_{{m}=1}^{N_T} c_{m}\,\ket{n_{1i}+{m},n_{2i}-2{m},n_{3i}+m}.
	\label{eq:TargetState}
\end{equation}
The Hilbert space described by the states $\ket{n_1,n_2,n_3}$, can be split into subspaces with fixed difference $n_1-n_3$, that contain the target state (\ref{eq:TargetState}).
Correlated processes are favoured by enforcing energy separation of these subspaces, i.e. we set  $\mu \gg J, U_{0}  , \mu_{2}$. We use both exact diagonalization and second-order perturbation theory in ${J}/{\mu}$ to analyze how a pair of bosons is transferred with high fidelity within one subspace. 

Two-atom and higher-order virtual processes allow to identify  an effective model for single pair transfer:
\begin{equation}
	H_{\text{eff,hop}}=J_{\text{eff}}\left[ \left(a_2^\dag \right)^2 a_1 a_3+ H.c. \right]~ , 
	\label{eq:EffectiveHoppingTerm}
\end{equation}

Consider the transfer of a single pair from the state $\ket{\psi_{i}}=\ket{n_1,n_2,n_3}$ to the state $\ket{\psi_{f}}=\ket{n_1+1,n_2-2,n_3+1}$. 
Following a second order effective hopping process, the intermediate virtual states are respectively $\ket{\psi_{l_1}}=\ket{n_1+1,n_2-1,n_3}$ and $\ket{\psi_{l_2}}=\ket{n_1,n_2-1,n_3+1}$.
By mean of a quasi-degenerate perturbation theory approach,
 the amplitude probability of the process is given by the sum over the contributions of all intermediate paths. We find that the effective hopping is given by
\begin{align}
\nonumber
	&J_{\text{eff}}^{II} = \\
	&-\frac{J^2 U_0  \left(\mu ^2+\mu _2^2+2 U_0 ^2+3 \mu _2 U_0 \right)}{\left(-\mu +\mu _2+U_0 \right) \left(-\mu +\mu _2+2 U_0 \right) \left(\mu +\mu _2+U_0 \right) \left(\mu +\mu _2+2 U_0 \right)}
	\label{eq:JeffIIStatic}
\end{align}
For $\mu_2=0$ and small interaction, $J_{\text{eff}}^{II}\simeq -J^2 U_0 /\mu^2 $. More generally, the second order hopping probability is zero for non-interacting particles $U_0 =0$, due to a destructive interference between the virtual paths such as:
\begin{eqnarray}
\nonumber
\ket{0,N,0}\,\rightarrow\,\ket{1,N-1,0}\,\rightarrow\,\ket{1,N-2,1}\\
\ket{0,N,0}\,\rightarrow\,\ket{0,N-1,1}\,\rightarrow\,\ket{1,N-2,1}
\end{eqnarray}.

\begin{figure}[htbp]
\includegraphics[width=0.6\columnwidth]{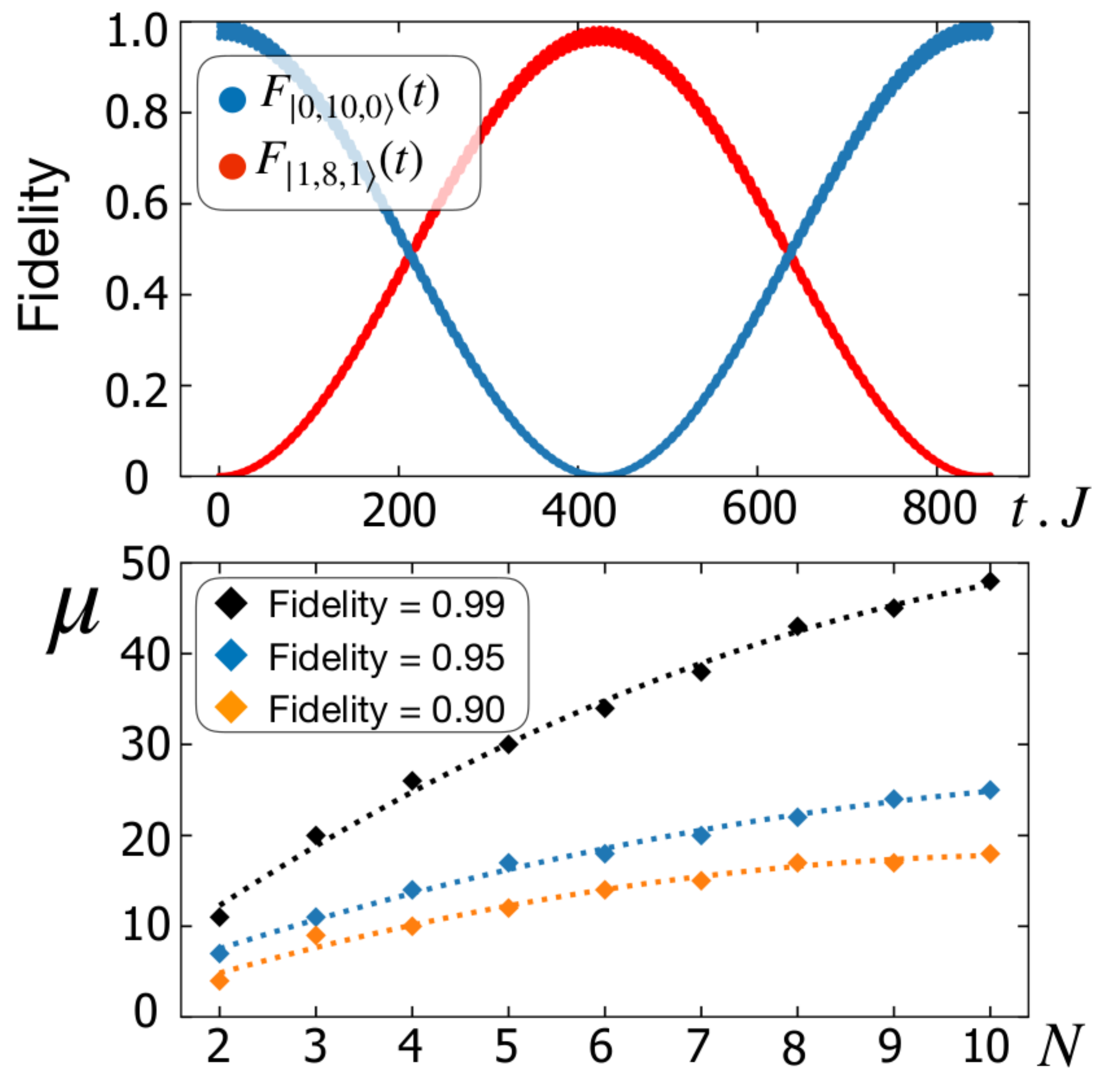}
\caption{(Top) Rabi-type oscillations of $F_{\ket{1,8,1}} = |\langle 1,8,1 | e^{-i Ht} |0,10,0 \rangle|^2$ and $F_{\ket{0,10,0}} = |\langle 0,10,0 |e^{-iHt}|0,10,0\rangle|^2$ (see text in Appendix A). Anticrossing makes a very long time scale emerge on  top of fast  single particle processes. Parameters are $\mu = 48$ , $U_0 = 0.9$,  $\mu_2 = 7.635$ (in $J$ units).
(Bottom) Dotted lines indicate the offset $\mu(N)$ required to achieve through tuning $\mu_2$ a given fidelity transfer from $|0,N,0\rangle$ to $|1,N-2,1\rangle$.}
\label{fig:SinglePairTransfer10a}
\end{figure}

Optimization of the resonance between two states differing by one nonlocal atom pair happens when two levels of the spectrum encounter an anticrossing. In the vicinity of the anticrossing, these states are little different from states $|0,N,0\rangle$ and $|1,N-2,1\rangle$. This generates a Rabi oscillation with nearly perfect fidelity on very long time scales (Fig. \ref{fig:SinglePairTransfer10a}).

Notice that in this correlated transport regime, the interaction $U_0$ 
must be large compared to the transfer rates $J_{\text{eff}}^{II}$.
As a result,  coherent combinations involving a large number of states in Eq.(\ref{eq:TargetState}) cannot be achieved with the present static protocol.  Single pair correlated transfer, though, is obtained with high fidelity by fine tuning the potential $\mu_2$ to compensate the total interaction energy.  It connects two states well-separated from the rest of the spectrum. High fidelity transfer can hold for any number of atoms $N=2N_T$ for large enough $\mu$, under the condition  that the manifolds of  states with the same $n_1-n_3$ are well-separated from each other (in order that second-order processes in ${J}/{\mu}$ is well-defined). Large $\mu$ leads to a nearly perfect resonance (Fig. \ref{fig:SinglePairTransfer10a}). Fig. \ref{fig:SinglePairTransfer10a} also 
shows the corresponding scaling $\mu(N_T, F)$ of the minimum $\mu$ necessary to obtain a given fidelity $F$. 
Readout of the resulting superposition can be achieved by time-of-flight (see Appendix E). 

\section{Appendix B: Driven interaction, Floquet expansion and the effective model.}
We consider a resonantly driven three-mode model ($\omega = 2\mu/\hbar$) with a strong offset between site 1 and 3.
\begin{align}
\nonumber
H&=-J\sum\limits_{i,j=1(i\neq j)}^{3}\left(a_j^\dag a_i +H.c.\right) + \\
&+ \frac{U_1}{2}\sin(\omega t)\sum\limits_{i=1}^{3} \hat{n}_i(\hat{n}_i-1) - \sum\limits_{i=1}^{3}\mu_i \hat{n}_i ~,
\label{eq:BoseHubbardShaken}
\end{align}
We apply the analysis carried out by Goldman et al.\cite{goldman2015periodically} (see also Ref. \onlinecite{eckardt_high-frequency_2015})  to obtain an effective model in the strongly driven regime.  Since  $U_{1}=K_{0}\omega$ where $K_{0} \sim 1$ and $\omega=2\mu/\hbar$, two terms diverge in the limit of $\omega\rightarrow\infty$. They can be eliminated away by a  unitary transformation. We first rewrite the Hamiltonian in a reference frame rotating with the driving term, as.
\begin{equation}
    \Tilde{H} = \mathcal{R}^{\dag}(t)H(t)\mathcal{R}(t) - i\mathcal{R}^{\dag}(t)\partial_{t}\mathcal{R}(t)
\end{equation}
with.
\begin{align}
    \mathcal{R}(t) &= -\exp{\left(i\int_{0}^{t}\mathcal{O}(\tau)\mathrm{d}\tau\right)} \\
		\mathcal{O}(t) &= \mu(\hat{n_{1}} - \hat{n_{3}}) + \frac{U_{1}}{2}\sin{(\omega t)}\sum_{j}\hat{n_{j}}(\hat{n_{j}} - 1)
\end{align}

The Hamiltonian takes the form :

\begin{eqnarray}
\nonumber
	\tilde{H}(t) &=& -J \Big( a_2^\dagger \exp\big[ i K_0\,(\hat{n}_2-\hat{n}_1) \sin\omega t - i \frac{\omega}{2} t\big ]\, a_1\\
	\nonumber
	 &+&a_2^\dagger \exp\big[i K_0\,(\hat{n}_2-\hat{n}_3) \sin\omega t + i \frac{\omega}{2} t \big]\, a_3\\
	\nonumber
	 &+&a_1^\dagger \exp\big[i K_0\,(\hat{n}_1-\hat{n}_3) \sin\omega t +  i \omega t \big] a_3 \Big)+H. c.~.\\
	 &=&\tilde{H}^{(21)}(t)+\tilde{H}^{(23)}(t)+\tilde{H}^{(13)}(t)
\end{eqnarray}
Following the high-frequency expansion, the effective Hamiltonian at order zero is given by :
\begin{equation}
	H_{eff}^{(0)} = \frac{1}{T}\int_{0}^{T}\Tilde{H}(t)\mathrm{d}t
\end{equation}

The term $\tilde{H}^{(13)}(t)$ obviously has a nonzero time average on a period $T=\frac{2\pi}{\omega}$, yielding the following term corresponding to transitions between sites $1$ and $3$:
\begin{equation}
H_{eff}^{(13)}=-J\,a_1^\dagger\,\mathcal{J}_1[K_0(\hat{n}_1-\hat{n}_3)]\,a_3\,+\,H.c.
\end{equation}
On the other hand, due to the half-frequency factor in the transitions from site $2$ to sites $1$ and $3$, the corresponding terms have a zero average on the period $2T$, which is the actual period of the transformed Hamiltonian $\tilde{H}$. Thus one needs to perform a Floquet expansion to first order in $\frac{1}{\omega}$, yielding terms of order $\frac{J^2}{\mu}$:

\begin{equation}
    H^{(21)}_{eff}+H^{(23)}_{eff} = \sum_{p > 0} \frac{1}{p\mu}\left[\Tilde{\mathcal{H}}^{p}_{12} + \Tilde{\mathcal{H}}^{p}_{23} \ ,\ \Tilde{\mathcal{H}}^{-p}_{12} + \Tilde{\mathcal{H}}^{-p}_{23} \right]
\end{equation}
where $\Tilde{H}^{p}_{ij}$ denotes the $p$-th harmonic of $\Tilde{H}_{ij}$.

The calculation of the harmonics and of the commutators is straightforward and yields Eq. 2
with the definitions:

\begin{widetext}
\begin{eqnarray}
\nonumber
\mathcal{L}(K_0,\hat{n},\hat{n}')\,&=&\,\sum_{m\geq0}\frac{1}{(2m+1)}\\
\nonumber
&&\Big(\mathcal{J}_{m + 1}[K_{0}(\hat{n}-\hat{n}' - 1)]\mathcal{J}_{-m}[K_{0}(\hat{n}-\hat{n}' - 3)] - \mathcal{J}_{-m}[K_{0}(\hat{n}-\hat{n}' - 1)]\mathcal{J}_{m + 1}[K_{0}(\hat{n}-\hat{n}' - 3)]\Big) \\
\nonumber
\mathcal{M}(K_0,\hat{n}_{i=1,2,3})\,&=&\,\sum_{m\geq0}\frac{1}{(2m+1)}\\
\nonumber
&&\Big(\mathcal{J}_{m}[K_{0}(\hat{n}_2-\hat{n}_1 - 1)]\mathcal{J}_{-m}[K_{0}(\hat{n}_2-\hat{n}_3 - 2)] - \mathcal{J}_{m}[K_{0}(\hat{n}_2-\hat{n}_1-2)]\mathcal{J}_{-m}[K_{0}(\hat{n}_2-\hat{n}_3-1)]\\
\nonumber
&+&\,\mathcal{J}_{m + 1}[K_{0}(\hat{n}_1-\hat{n}_2 - 1)]\mathcal{J}_{-(m+1)}[K_{0}(\hat{n}_3-\hat{n}_2 - 2)] - \mathcal{J}_{m+1}[K_{0}(\hat{n}_1-\hat{n}_2-2)]\mathcal{J}_{-(m+1)}[K_{0}(\hat{n}_3-\hat{n}_2-2)]\Big) \\
\nonumber
\mathcal{N}(K_0,\hat{n}_{i=1,2,3})\,&=&\,\sum_{m\geq0}\frac{1}{(2m+1)}\\
\nonumber
&&\Big[\Big(\mathcal{J}_{m+1}[K_{0}(\hat{n}_1-\hat{n}_2- 1)]\mathcal{J}_{-m}[K_{0}(\hat{n}_2-\hat{n}_3)] \,- \,\mathcal{J}_{-m}[K_{0}(\hat{n}_1-\hat{n}_2)]\mathcal{J}_{m+1}[K_{0}(\hat{n}_2-\hat{n}_3+1)]\Big)\,(\hat{n}_2+1) \\
\nonumber
&+&\,\mathcal{J}_{-m}[K_{0}(\hat{n}_1-\hat{n}_2 + 1)]\mathcal{J}_{m+1}[K_{0}(\hat{n}_2-\hat{n}_3)] \,- \mathcal{J}_{m+1}[K_{0}(\hat{n}_1-\hat{n}_2)]\mathcal{J}_{-m}[K_{0}(\hat{n}_2-\hat{n}_3-1)]\,\hat{n}_2\Big)\Big]\\
\mathcal{P}(K_0,\hat{n},\hat{n}')\,&=&\,\sum_{m\geq0}\frac{1}{(2m+1)}\Big(\mathcal{J}_{m }[K_{0}(\hat{n}-\hat{n}' - 1)]^{2}- \mathcal{J}_{m + 1}[K_{0}(\hat{n}-\hat{n}' - 1)]^{2}\}\Big)
\label{eq:BesselSums}
\end{eqnarray}
\end{widetext}

This effective Hamiltonian can be further simplified in the limit of large interaction $K_0\gg1$, where the asymptotic expansion holds:

\begin{equation}
J_{n}(x) \sim \sqrt{\frac{2}{\pi x}}\cos{(x - \frac{n\pi}{2} - \frac{\pi}{4})}
\end{equation}
yielding after a few steps the following form of the second order terms in $J/\mu$:

\begin{widetext}
\begin{equation}
H_{eff}^{Asympt,(2)}\,=\,\left[\mathcal{H}^{(12)}(a^{\dagger}_1\,a_2)^{2}\,+\,\mathcal{H}^{(23)}(a^{\dagger}_2\,a_3)^{2}
\,+\,\mathcal{H}^{(123)}(a^{\dagger}_2)^{2}\,a_1\,a_3\right]\,+\,H. c.\,+\,\mathcal{H}_{pot}^{(123)}
\end{equation}
with:

\begin{eqnarray}
\nonumber
\mathcal{H}^{12}&=&\frac{J^2}{2K_0\mu}\left[\delta_{\hat{n}_{12},2}\,\cos(2K_0) \,+\,(1-\delta_{\hat{n}_{12},1})(1-\delta_{\hat{n}_{12},2})(1-\delta_{\hat{n}_{12},3})\,\sin(2K_0)\right]\frac{1}{\sqrt{|\hat{n}_{12} - 1||\hat{n}_{12} - 3|}}\\
\nonumber
&+& \frac{J^2}{\mu}\left[\delta_{\hat{n}_{12},1}+\delta_{\hat{n}_{12},3}\right]\,\mathcal{J}_{1}(2K_0)\\
\nonumber 
\mathcal{H}^{23}&=&\frac{J^2}{2K_0\mu}\left[\delta_{\hat{n}_{23},2}\,\cos(2K_0) \,+\,(1-\delta_{\hat{n}_{23},1})(1-\delta_{\hat{n}_{23},2})(1-\delta_{\hat{n}_{23},3})\,\sin(2K_0)\right]\frac{1}{\sqrt{|\hat{n}_{23} - 1||\hat{n}_{23} - 3|}}\\
\nonumber
&+&\frac{J^2}{\mu}\left[\delta_{\hat{n}_{23},1}+\delta_{\hat{n}_{23},3}\right]\,\mathcal{J}_{1}(2K_0)\\
\nonumber
\mathcal{H}^{123}&=& \frac{J^2}{2K_{0}\mu}\left[\frac{(1-\delta_{\hat{n}_{32},1})(1-\delta_{\hat{n}_{12},2})}{\sqrt{|\hat{n}_{12} - 1||\hat{n}_{32}- 2|}}\,\cos{\left(K_{0}(\hat{n}_{13} -1)\right)}\,-\,\frac{(1-\delta_{\hat{n}_{32},2})(1-\delta_{\hat{n}_{12},1})}{\sqrt{|\hat{n}_{12} - 2||\hat{n}_{32}- 1|}}\,\cos{\left(K_{0}(\hat{n}_{13}+1)\right)}\right]\\
\nonumber
&+&\,\left[\delta_{\hat{n}_{12},2}\mathcal{J}_0\left(K_0(\hat{n}_{32} - 1)\right)\,+\,\delta_{\hat{n}_{32},1}\mathcal{J}_0\left(K_0(\hat{n}_{12} - 2)\right)\,-\,\delta_{\hat{n}_{12},1}\mathcal{J}_0\left(K_0(\hat{n}_{32} - 2)\right)\,-\,\delta_{\hat{n}_{32},2}\mathcal{J}_0\left(K_0(\hat{n}_{12} - 1)\right)\right]\\
\nonumber
\mathcal{H}_{pot}^{123}&=&\frac{J^2}{2\mu}\left[\mathrm{sinc}{\left(2K_{0}(\hat{n}_{21} - 1)\right)}\,\hat{n}_{1}\,(\hat{n}_{2} + 1) - \mathrm{sinc}{\left(2K_{0}(\hat{n}_{21} + 1)\right)}\,\hat{n}_{2}\,(\hat{n}_{1} + 1)\right.\\
    &+& \left. \mathrm{sinc}{\left(2K_{0}(\hat{n}_{32} - 1)\right)}\,\hat{n}_{2}\,(\hat{n}_{3} + 1) - \mathrm{sinc}{\left(2K_{0}(\hat{n}_{32} + 1)\right)}\,\hat{n}_{3}\,(\hat{n}_{2} + 1)\right] \,
\label{asymptotic}
 \end{eqnarray}
\end{widetext}
where $\mathrm{sinc}{(x)}=\frac{\sin{x}}{x}$. Inspection of the successive terms helps to understand several features of the exact numerical solution. First, the term $\mathcal{H}^{123}$ vanishes when the wells $1,3$ have the same occupation. Second, terms $\mathcal{H}^{12}$, $\mathcal{H}^{23}$ vanish when $K_{0,n}=\frac{n\pi}{2}$, therefore Coherent Destruction of Tunneling (CDT) is obtained starting from state $\ket{0,N,0}$ with $K_0$ close to those $K_{0,n}$ values. Despite the approximation contained in this asymptotic expansion, it explains very well the map of Fig. \ref{fig:Exact_vs_Eff}, middle panel.

 \begin{figure}[htb]
  \centering
 	\includegraphics[scale=0.3]{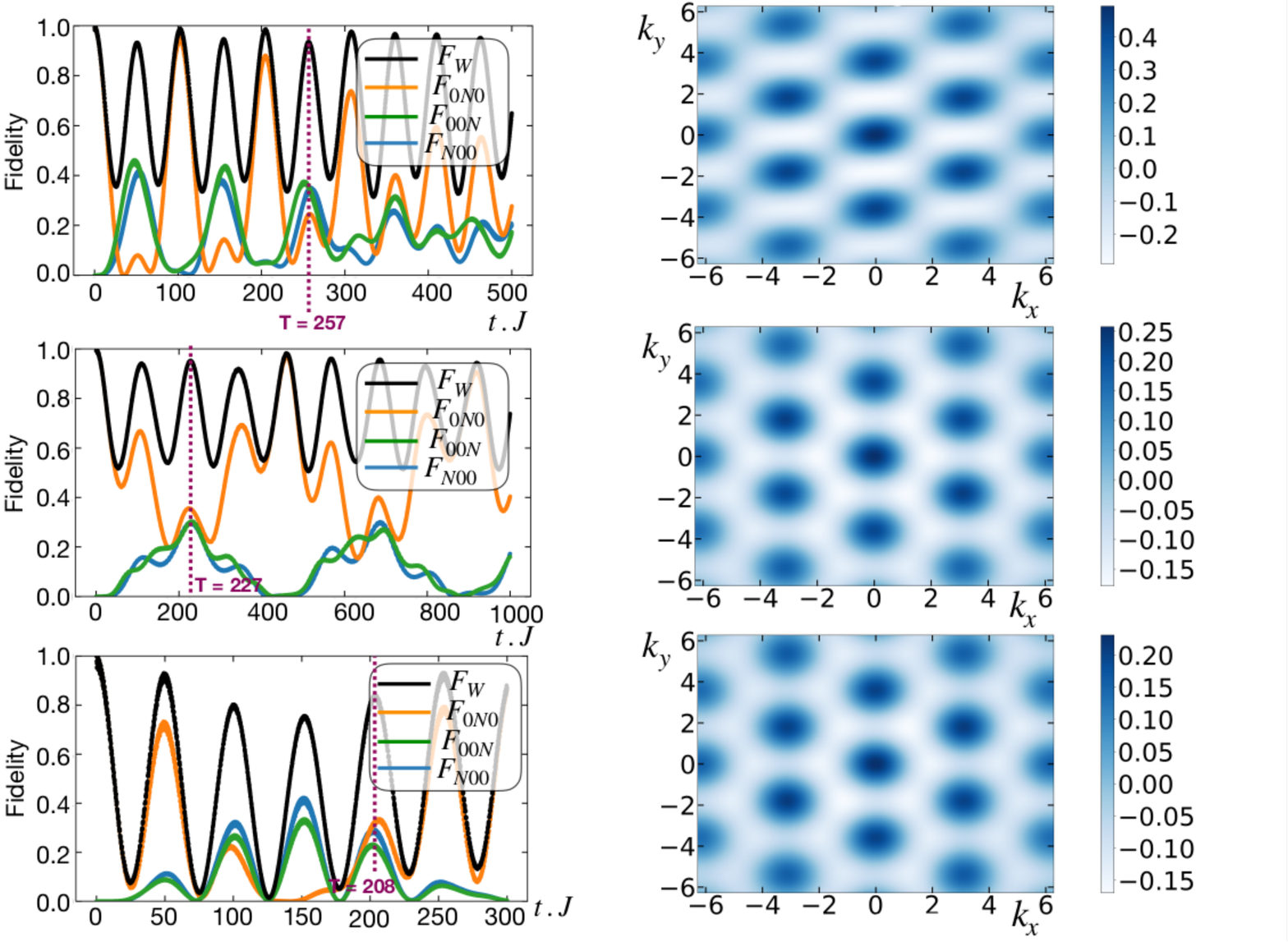}
 	\caption{\small{Trajectories and readout maps for (from top to bottom) $N=4,\, K_0=2.4,\, \mu=15$; $N=6,\, K_0=2.7,\, \mu=19$; and $N=10,\, K_0=1.16, \,\mu=11.7$, starting at $t=0$ from state $|0N0\rangle$: (left panels) Fidelities for states $|0N0\rangle,|N00\rangle,|00N\rangle$ and their sum $F_W$; (right panels) two-atom correlation $\sigma({\bf k},{\bf -k})$ (see Section V and Appendix E) for a W-type state realized at times indicated by a vertical dotted line on the left panels.} }
 	\label{fig:Supplemental_Map_N}
 \end{figure}

\section{Appendix C: Additional data with a driven interaction.}
 We present here some wavefunction trajectories for other even atom numbers. For $N=4, 6,10$ one finds the same trends as in Fig. \ref{fig:Exact_vs_Eff}, bottom panel: tuning the interaction $K_0$ allows to periodically achieve a high-fidelity superposition of states $|0N0\rangle,|N00\rangle,|00N\rangle$, in particular W-type states (Figure \ref{fig:Supplemental_Map_N}). Notice the nearly periodical appearance of such states, with a very long period (several $100 J^{-1}$) related to anticrossings in the Floquet pseudo-energy spectrum. The larger $N$, the finer the tuning of $K_0$ necessary to obtain high fidelities. 

\begin{figure}[htb]
  \centering
 	\includegraphics[scale=0.35]{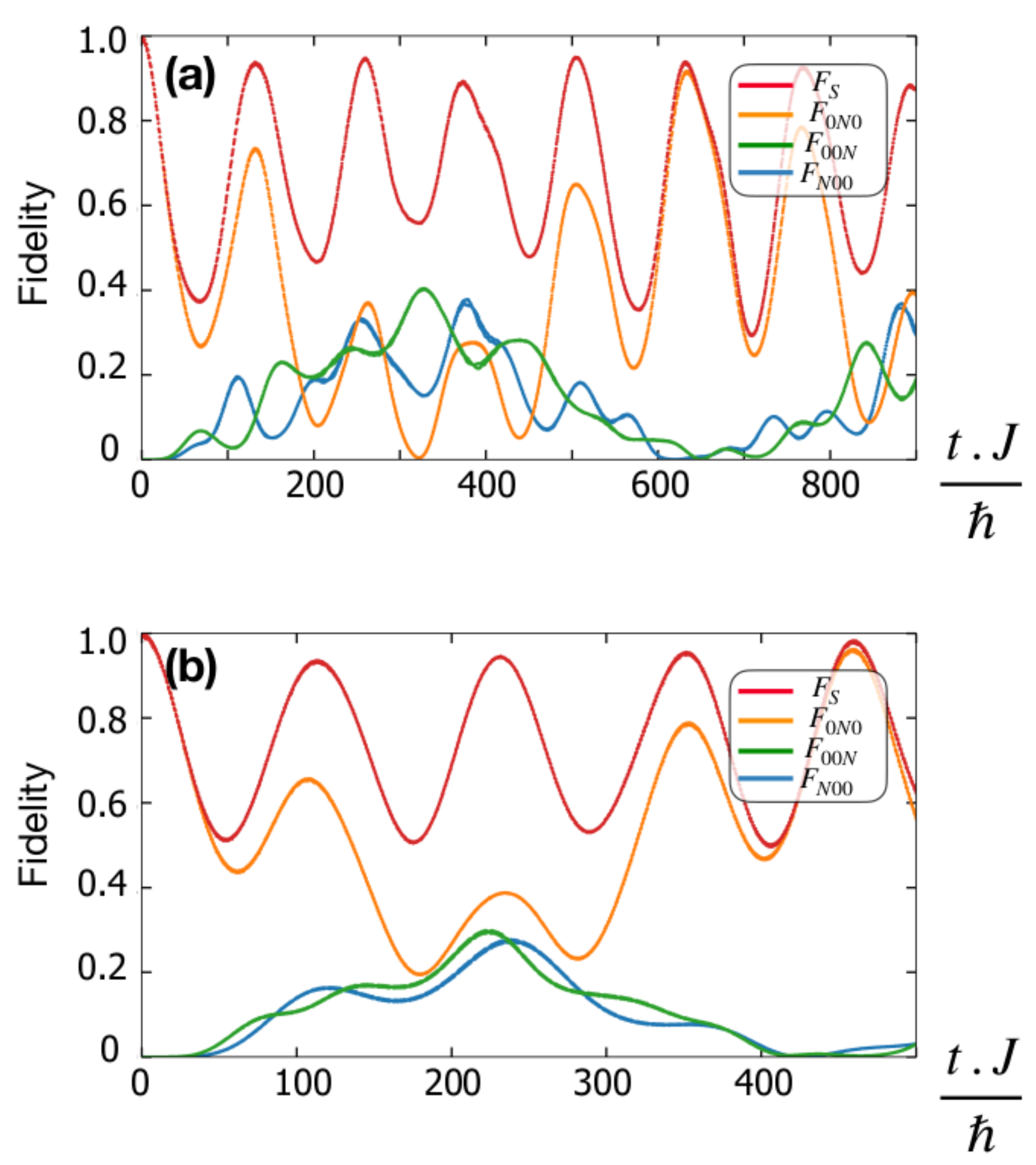}
 	\caption{\small{Trajectories for an asymmetric 3-site ring (top) $N=6,\, \mu=17, \,K_0=2.65,\, J_{12}=0.9,\, J_{23}=1.1,\,J_{13}=1$ and (bottom) $N=6,\, \mu=17,\, K_0=2.7, \,J_{12}=J_{13}=1,\, J_{13}=0.9$.} }
 	\label{fig:asymmetries}
 \end{figure}

We have also tested the effect of coupling asymmetries in the ring geometry ($M=3$). Writing the inter-site coupling as $H_J=\sum_{i\neq j} J_{ij}\,a^{\dagger}_ja_i$, 3-mode N00N states can still be generated with  moderate asymmetry $J_{12}\neq J_{32}$ (Fig. \ref{fig:asymmetries}). On the other hand, keeping $J_{12}= J_{32}$ but taking $J_{13}\neq J_{12}$, N00N states also occur unless $J_{13}$ is too small. 

\begin{figure}[htb]
  \centering
 	\includegraphics[scale=0.5]{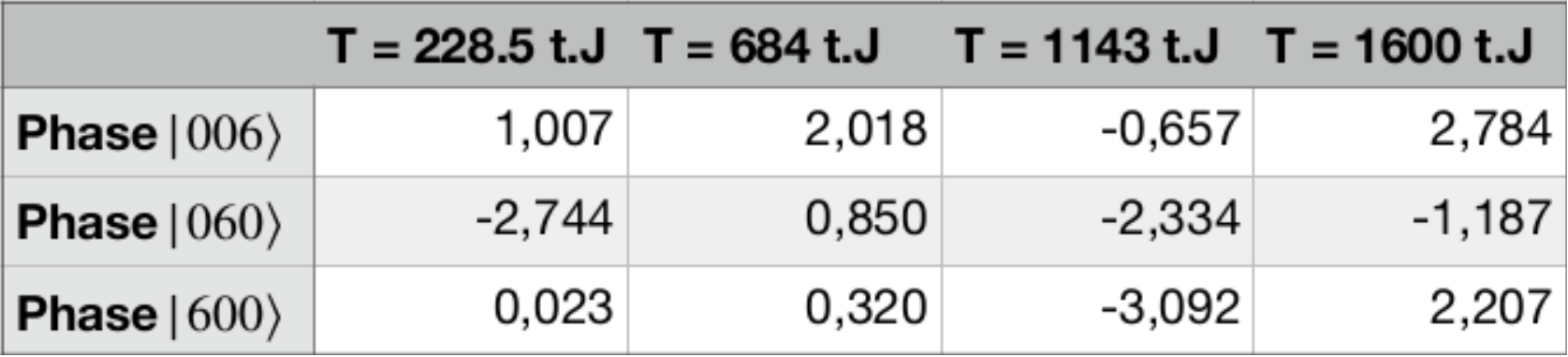}
 	\caption{\small{Phases (in radians) of the coefficients of states $\ket{006},\ket{060},\ket{600}$ for different W-states obtained successively in time.} }
 	\label{table_phases}
 \end{figure}

An important point is the coherence of the multi-mode N00N states. First, as resulting from the unitary evolution of a pure state $\ket{0N0}$, they are by construction pure states. We do not discuss here the possible causes of decoherence due to external fluctuations. Let us instead focus on the relative phases of the states forming such states i.e. the phases of the coefficients $a_i(t)$ of a state:
\begin{equation}
\Psi_S(t_k)\,=\,\sum_i\,a_i(t_k) \, \ket{00..N_i..00}
\end{equation}
We find that the $a_i$'s are in general complex numbers, and their phases, fixed for a given time, are not correlated from an occurrence time $t_k$ to another $t_l$. This comes from the nontrivial interferences occurring in the many-body wavefunction.
As an example, Figure \ref{table_phases} gives the phases of successive N00N states components.

\section{Appendix D: Formation of three-mode N00N states by the emergence of very low energy scales.}
Let us show how our protocol defines a hierarchy of energy scales, down to very low ones where N00N states form by resonance of very few low-energy Floquet states. The analysis is performed on the effective Hamiltonian, which provides a very accurate qualitative and almost quantitative description of the motion, averaged on the period $\hbar/\mu$.  Compared to the exact dynamics, the one led by $H_{eff}$ misses the fast motion (on times $\hbar/\mu$) and yields some irrelevant drift at longer scales, without changing the main conclusions. The scheme will be illustrated on $N=2,4,6$ examples with three wells. It generalizes to any even $N$.

The Floquet averaging leaves two kinds of terms in $H_{eff}$. Direct averaging of the transitions between wells $1$ and $3$ yields a dressed single-atom hopping term of order $J \ll \mu$. This is, hierarchically speaking, the second energy scale in the problem. The third energy scale is $J^2/\mu \ll J$, it governs three kinds of terms, given by series of products of Bessel functions: (i) atom pair hoppings from $2$ to $1$, from $2$ to $3$, and from $2$ to ($1,3$) simultaneously, as a nonlocal pair; (ii) corrections to the hopping between $1$ and $3$; (iii) potential terms. Notice that when the occupations $n_1$ and $n_3$ are equal, pair hopping from $2$ to ($1,3$) is forbidden by a cancellation of terms in the Bessel series. This situation reminds that encountered in the undriven case with a constant repulsive interaction (Appendix A), where such transitions are found only if the repulsive interaction is nonzero. In the driven case, the average interaction is in fact zero.

One also notices that within the Fock space $\ket{n_1,n_2,n_3}$, $H_{eff}$ splits into two uncoupled blocks, one made with  even $n_2$ and the other made with odd $n_2$. Starting with the initial state $\ket{0N0}$ with N even, the dynamics restricts to states $\ket{n_1,n_2,n_3}$ with $n_2$ even but $n_1, n_3$ can still be even or odd. This leaves $4$ states over $6$ for $N=2$, $9$ states over $15$ for $N=4$, $16$ states over $28$ for $N=6$, more generally $(\frac{N}{2}+1)^2$ states.

\subsection{The limit of infinite $\mu$.}
Let us treat $H_{eff}$ by solving first the infinite-$\mu$ limit, then in perturbations in $J/\mu$. Dropping all the $J^2/\mu$ terms yields $(1+\frac{N}{2})$ blocks of states where $n_2$ is constant (and even), connected by $1\leftrightarrow3$ hopping. Block $n$ ($n=0...\frac{N}{2}$) is formed by $(2n+1)$ Fock states:

\begin{equation}
\ket{2n,N-2n,0},\;\ket{2n-1,N-2n,1}...\ket{0,N-2n,2n}
\end{equation}

This subspace is represented by a symmetric tridiagonal matrix with all diagonal terms being zero, and successive couplings $T_{n,1},T_{n,2},..T_{n,2n}$, which are of order $J$ and obey $T_{n,m}=-T_{n,2n-m+1}$ (change of sign by interchanging $n_1$ and $n_3$), as can be understood by inspection of the Bessel functions in Eq. (\ref{eq:BesselSums}). As a result, the matrix is simplified by taking the combinations:

\begin{equation}
|\psi_{n,m,\pm}\rangle=2^{-1/2}(\ket{2n-m,N-2n,m}\pm\ket{m,N-2n,2n-m})
\label{combinations}
\end{equation}

The dimension of this matrix is $2n+1$ and its spectrum consists of n pairs of  states with energies $\pm E^{(0)}_{n,\kappa}$ ($\kappa=1,..n$) of order $J$, {\it and one state at $E=0$}. To show this, one considers the two cases, $n$ even and $n$ odd and observe that this matrix further splits into two blocks. In the first one $|\Psi_{n,0,+}\rangle$ is coupled to $|\Psi_{n,1,-}\rangle$ and successively to $|\Psi_{n,n/2,+}\rangle$ if $n$ is odd, and up to the "central" state $\ket{n,N-2n,n}$ if $n$ is even. The second point is that such tridiagonal matrices possess a zero eigenvalue if their dimension is odd. One checks that this is always the case with the matrix containing the state $|\Psi_{n,0,+}\rangle=2^{-1/2}(\ket{2n,N-2n,0}+\ket{0,N-2n,2n})$. For $n=N/2$ this is nothing but the N00N state $2^{-1/2}(\ket{N00}+\ket{00N})$ built on wells $1,3$. Moreover, the zero-energy state of each block is obtained by successively eliminating all odd combinations, leaving only "even" states $\ket{n-2p,N-2n,2p}$. Therefore, {\it all zero-energy states in the infinite $\mu$ limit are made exclusively with even occupations of all three sites $1,2,3$}, spanning a reduced Hilbert space of dimension $(N+2)(N+4)/8$ instead of $(N+1)(N+2)/2$ in the full space. We show in the following that the essential of the dynamics takes place within this restricted boson pair Fock space. 

The $n$-th $E=0$ state can be formally written as:

\begin{eqnarray}
|\psi^{(0)}_{n}\rangle&=&\alpha_0(\ket{2n,N-2n,0}+\ket{0,N-2n,2n})\\
\nonumber &+&\alpha_1(\ket{2n-2,N-2n,2}+\ket{2,N-2n,2n-2}) +..
\end{eqnarray}
The $\alpha$'s are coefficients of order $1$. One has $|\psi^{(0)}_{0}\rangle=\ket{0N0}$, and $\ket{\psi^{(0)}_{N/2}}$ contains the $(1,3)$ N00N state. Those remain to be connected to form the W-state.

Gathering all blocks thus yields for the infinite-$\mu$ limit $(1+\frac{N}{2})$ zero-energy states and $\frac{N}{4}(1+\frac{N}{2})$ pairs of high-energy states with opposite energies of order $J$. The latter can be written with generality:

\begin{equation}
|\psi^{(H)}_{n,\kappa}\rangle=\sum_{m,\pm}\beta_{n,\kappa,m,\pm}\,|\Psi_{n,m,\pm}\rangle
\end{equation}
with coefficients of order $1$.

\subsection{Perturbation in $J^2/\mu$.}
Let us now sketch the perturbative effect of all pair and potential terms. First, it raises the degeneracy of the $(1+\frac{N}{2})$ zero-energy states, yielding $(1+\frac{N}{2})$ states whose energies are of order $J^2/\mu$. They are separated from the remaining states that stay at order $J$. Without details, one can write these low-energy states as:

\begin{equation}
|\Psi^{(L)}_{\zeta}\rangle=\sum^{N/2}_{n=0}\,a_{\zeta,n}\ket{\Psi^{(0)}_{n}}
\end{equation}
with coefficients of order $1$ in general (except close to a CDT, see below), and energies $\varepsilon_{\zeta}$ of order $J^2/\mu$. On the other hand, the high-energy states mix together to form states $\ket{\Psi^{(H)}_{\kappa}}$ with energies $E_{\kappa}$ of order $J$.

The dynamics from the initial state $\ket{0N0}$ is obtained by expanding this state at $t=0$ in the basis ${\ket{\Psi^{(L)}_{\zeta}},\ket{\Psi^{(H)}_{\kappa}}}$, which yields at time $t$:

\begin{equation}
\ket{\Psi(t)}=\sum_{\zeta}\,a_{\zeta}e^{i \varepsilon_{\zeta}t}\ket{\Psi^{(L)}_{\zeta}}\,+\,\sum_{\kappa}b_{\kappa}e^{ i E_{\kappa}t}\ket{\Psi^{(H)}_{\kappa}}
\end{equation}

The first part of the sum with coefficients $a_{\zeta}$ of order $1$ contains only even occupation states, while the second part contains all states, but with small coefficients $b_{\kappa,n,\pm}$ of order $J/\mu$. Let us consider the transition probability to state $2^{-1/2}(\ket{N00}+\ket{00N})$, mainly contained in the $\ket{\Psi^{(L)}_{\zeta}}$'s, with a small component in the $\ket{\Psi^{((H)}_{\kappa}}$. It is essentially given by:

\begin{equation}
P_{N00+00N}(t)\,=\,\sum_{\zeta,\zeta'}c_{\zeta,\zeta'}\cos(\varepsilon_{\zeta}-\varepsilon_{\zeta'})t
\end{equation}
with $c_{\zeta,\zeta'}$ of order $1$, plus terms of order $(J/\mu)^2$, and it oscillates slowly. This formal calculation shows that the probability to form the state $2^{-1/2}(\ket{N00}+\ket{00N})$ is large in the course of time. Moreover, corrections will also weakly couple to state $2^{-1/2}(\ket{N00}-\ket{00N})$ (participating to the high-energy states), thus explaining why in the trajectories, the fidelities for states $\ket{N00}$ and $\ket{00N}$ are quite similar but not identical (small corrections produce drift at very long time scales). 

The conclusion of this perturbative analysis is that the interesting dynamics merely develops in a low-energy sector emerging from slightly perturbed degenerate states, with one in each block with constant $n_2$. The number of these states is $1+(N/2)$. Despite one works in a three-mode system, it is similar to the number of states of a two-mode system, with in addition a perfect pair correlation making a huge reduction of the chain of states connecting $\ket{0N0}$ to states $\ket{N00}$, $\ket{00N}$.

\begin{figure}[htb]
  \centering
 	\includegraphics[scale=0.13]{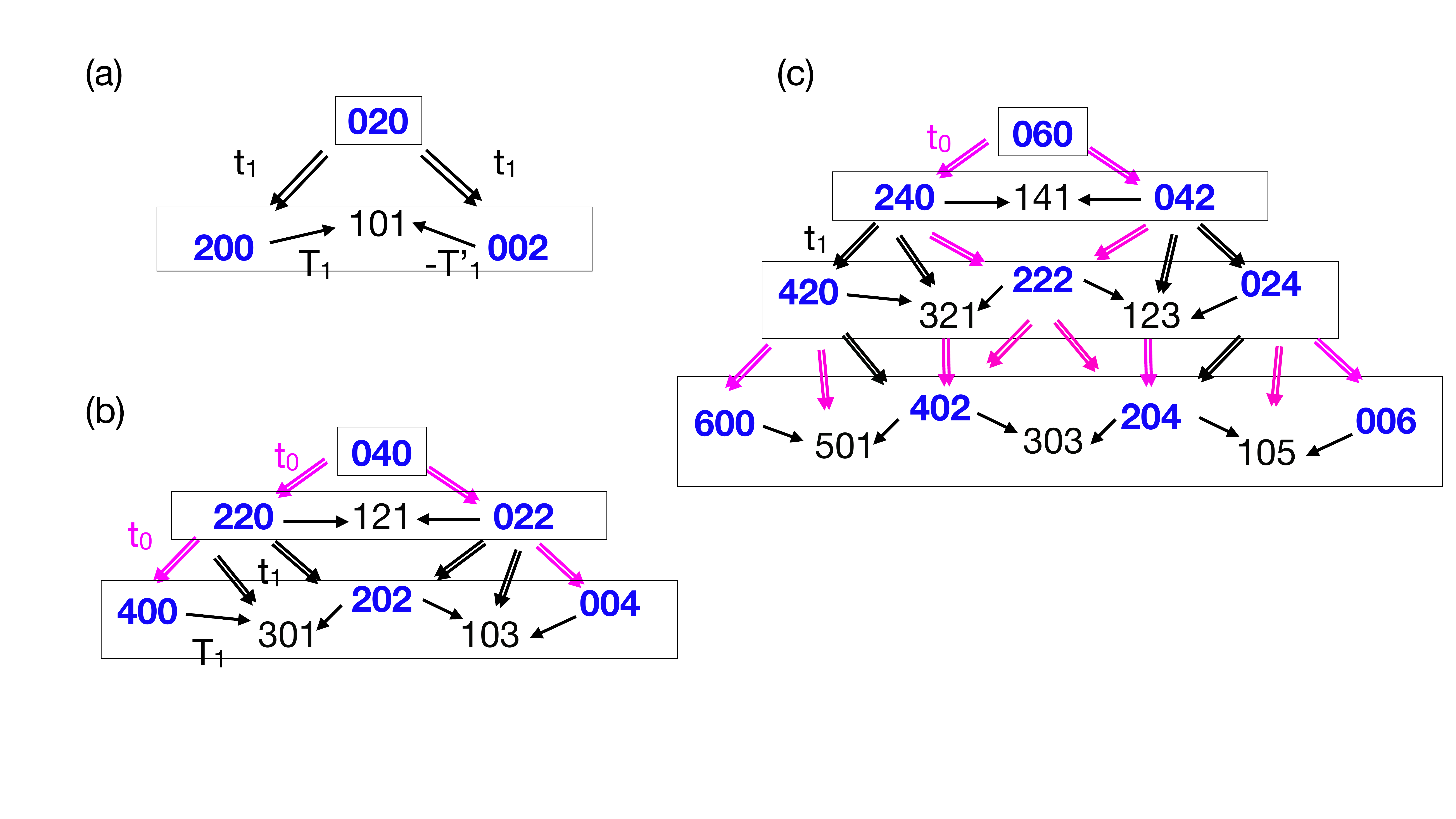}
 	\caption{\small{Fock states for $N=2\,(a)$, $N=4\,(b)$, $N=6\,(c)$. At infinite $\mu$, rectangles delimit decoupled blocks, each containing one zero-energy eigenstate of $H_{eff}$. Thin black arrows denote single-atom transitions (couplings $T_1=\mathcal{O}(J)$) with $T_1=T'_1$ if $\mu=\infty$), double black arrows denote pair transitions (couplings $t_1=\mathcal{O}(J^2/\mu)$) and double red arrows pair transitions vanishing at CDT ($K_0=n\pi/2)$, having couplings $t_0=\mathcal{O}((J^2/\mu)(\delta K_0/K_0))$ (represented only for $N=4,6$).} }
 	\label{configurations}
 \end{figure}

\begin{figure*}[htb]
  \centering
 	\includegraphics[scale=0.45]{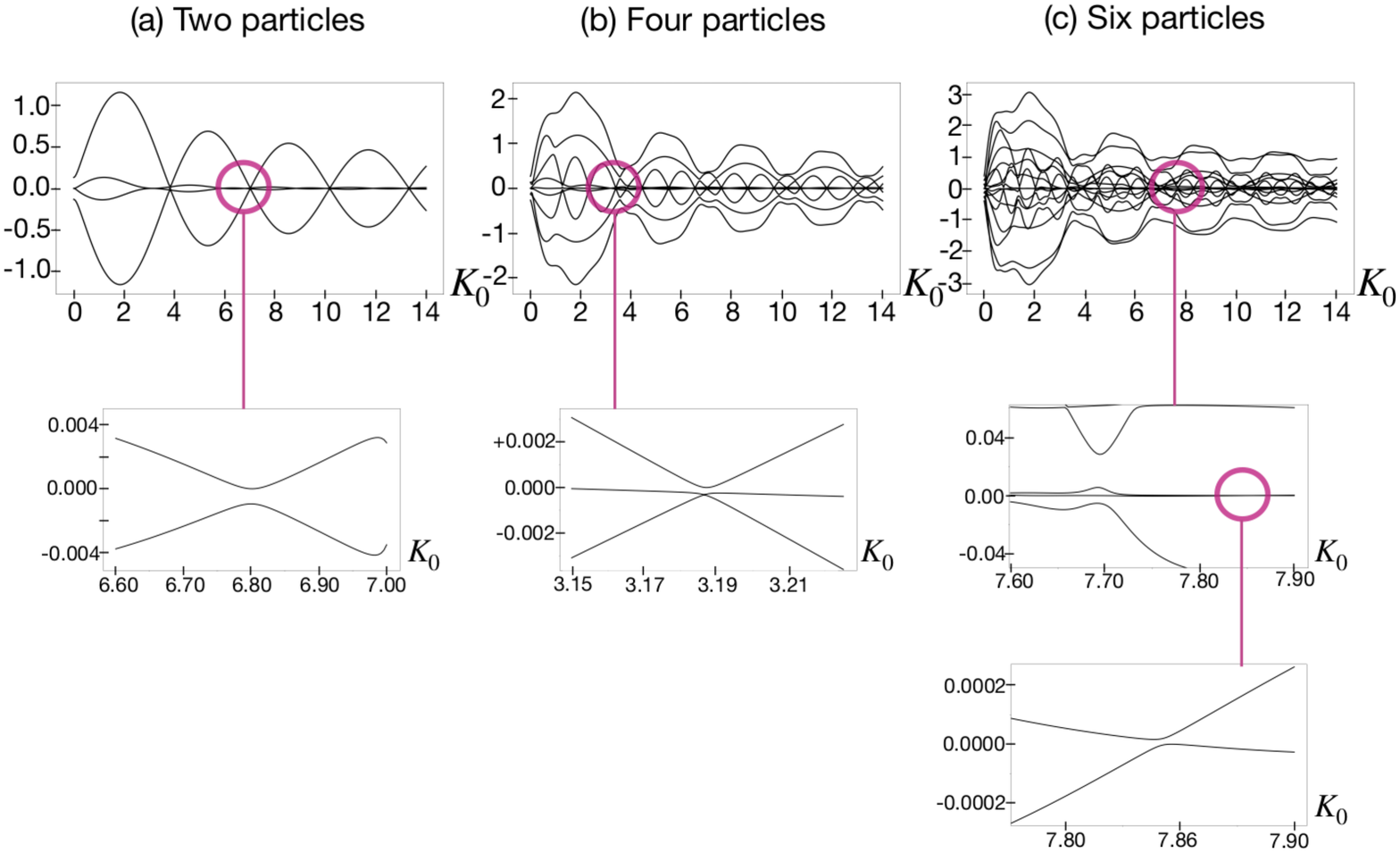}
 	\caption{\small{Spectrum of the effective Hamiltonian for $N=2\,(a)$, $N=4\,(b)$, $N=6\,(c)$ at different scales, showing the low-energy states emerging in the center of the spectrum: two states for $N=2$, three states for $N=4$, four states for $N=6$ of which two are only distinguished at very low scale. Zooming is made in the region of a CDT.} }
 	\label{Spectra_Heff}
 \end{figure*}

\subsection{Proximity to CDT.}
The latter reasoning holds for any value of the interaction drive $K_0$, but it does not yet explain why among the $(N+2)(N+4)/8$ pair states, excellent W-states made of well-balanced superpositions of $\ket{0N0}$, $\ket{N00}$ and $\ket{00N}$ can form for some values of $K_0$. For values of $N$ larger than $4$, those can indeed be found in the vicinity of CDT regimes where the system is merely blocked in the initial state. The asymptotic expansion shows that this happens when $K_0=n\pi/2$. CDT due to vanishing of Bessel function dressing has been found in the past for coupled wells \cite{gong_many-body_2009}, and Watanabe \cite{watanabe_floquet_2012} noticed the proximity of CDT to good N00N states generated in such a simpler system. We hereafter give an explanation in our more general case.

The low-energy states connecting $\ket{0N0}$ to states $\ket{N00}$, $\ket{00N}$ are successively $\ket{\Psi^{(L)}_1},\ket{\Psi^{(L)}_2}...\ket{\Psi^{(L)}_{N/2}}$. Let us call their mutual couplings (of order $J^2/\mu$) $t^{(L)}_1,t^{(L)}_2...,t^{(L)}_{N/2}$, that result from the amplitudes of the terms in $H_{eff}$ transferring pairs one by one from, say, well $2$ to well $1$. The asymptotic form of these terms does not always vanish at CDT: it does so if $n_1-n_2 < 1$ or if $n_1-n_2 > 3$, numbers being taken in the final state of each transition (see Eq. \ref{asymptotic}). As a result, $t_i$'s vanish except for $n_1-n_2 =2$, which happens once in the connecting chain if $N/2$ is odd. It it is even, a few other couplings to even state will be nonzero (see Fig. ). In practice, working close to a CDT point and expanding the cardinal sine function with $K_0=n\pi/2+\delta K_0$, most pair couplings $t_i$'s are of order of the very low scale $t^{(vL)}=(J^2/\mu)(\delta K_0/K_0)$, and a few ones stay of order $J^2/\mu$. 

At this stage one can repeat the scheme already used in front of two different energy scales: diagonalize the matrix at CDT, followed by a perturbative expansion. At CDT one finds that states $\ket{0N0},\ket{N00},\ket{00N}$ are disconnected, as well as the states of the other first blocks $n<N/4$, which happens only for $N \geq 8$. Other states lie at energies of order $J^2/\mu$ due to their nonvanishing couplings at CDT. Then, deviating slightly from CDT, transitions from $\ket{0N0}$ to $\ket{N00}$ or to $\ket{00N}$ involve perturbative couplings trough those states, with amplitudes $t^{(vL)}$. This eventually generates an effective coupling between N00N components, of order $(t^{(vL)})^2/(J^2/\mu)\sim(J^2/\mu)(\delta K_0/K_0)^2$. The latter fixes the typical recurrence frequency $\omega_{rec}$ of W-states.

At these energy and time scales we are left with the three N00N state components, plus a few even states sitting in the first blocks (zero for N=2-6, two for N=8, 10 etc...). Eventually, the good fidelity of the three-mode N00N state at some precise times relies on cancelling the probabilities of those "parasitic" states by interferences, which is quite easy to achieve when their number is small. To summarize, our hierarchical scheme - a kind of ad hoc renormalization - has successively eliminated:

(i) Odd $n_2$ states by approximating the exact dynamics by that of $H_{eff}$ (at order $J/\mu$).

(ii) Odd $n_1,n_3$ states at order $J/\mu$.

(iii) Close to CDT, most pair states except $\ket{0N0},\ket{N00},\ket{00N}$ and very few others. 

\noindent
The hierarchy of energies thus generated is:
\begin{eqnarray}
\mu\;\gg\;J\;\gg\;J^2/\mu\;\gg\;t^{(vL)}&=&J^2/\mu\,(\delta K_0/K_0)\\
\nonumber
&\gg&\;\omega_{rec}=J^2/\mu\,(\delta K_0/K_0)^2
\end{eqnarray}

\subsection{Examples: $N=2$, $N=4$, $N=6$.} 
For $N=2$ there are only two degenerate states at infinite $\mu$ : $\ket{020}$ and $2^{-1/2}(\ket{200}+\ket{002})$. The only remaining even $n_2$ state is $\ket{101}$, which will happen only at order $(J/\mu)^2$. Therefore, the dynamics starting from $\ket{020}$ reaches excellent W-states of two bosons, whatever $K_0$ and at any time. Nearly balanced W-states can be found at regular times. In this simple case there is no need for CDT. The reduction of the Hilbert space to a few atom pair states suffices, which shows the power of our scheme based on strong offset and resonant drive, making atom pair correlations. 

For $N=4$, the three infinite-$\mu$ degenerate states are $\ket{040}$, $2^{-1/2}(\ket{220}+\ket{022})$ and $\alpha(\ket{400}+\ket{004})+\beta\ket{202})$ with $\alpha, \beta$ numbers of order $1$. Fig. \ref{configurations} shows in red the couplings that vanish at CDT, and in black the small (double arrow, order $J^2/\mu$) and the large (simple arrow, order $J$) couplings that are insensitive to CDT. Close to CDT, the resulting dynamics essentially involves states $\ket{040},\ket{400},\ket{004}$, the others being split at higher energy. Actually, the small size of the pair Hilbert space makes possible to find N00N states even far from CDT values. This shows that our protocol is very robust already to maximally entangle $2$ or $4$ particles in three modes, which is a nontrivial achievement.

The case $N=6$ illustrates well the above discussion. Four blocks yield four degenerate states at infinite $\mu$. Among them, the initial state $\ket{060}$ and the combination $2^{-1/2}(\ket{600}+\ket{006})$ are specially favoured because they are very weakly coupled to other states close to CDT. This for instance generates N00N states on time scales $\sim 10^4 \hbar/J$ for $K_0\sim7.73$, close to  $5\pi/2$. 

Fig. \ref{Spectra_Heff} shows the spectrum of $H_{eff}$ for $N=2,4,6$, first in a large $K_0$ range: one clearly sees a quasi-periodicity and a decreasing envelope characteristic of the Bessel functions forming the various components of the Hamiltonian. Second, in a more restricted $K_0$ range containing a CDT, the low-energy states are plotted. One more zoom is made in the $N=6$ case to show the very low energy states and the anticrossing at the CDT. Notice that there is no strong anomaly in the spectrum at an optimum $K_0$ value for N00N states ($K_0=7.73$ for $N=6$).

\section{Appendix E: Readout maps, details and exploitation.}
The two states $|\Psi_{W}\rangle$ and $|\Psi_{pair}\rangle$ that we have shown to be generated in our system are superpositions of states that are connected by at least one pair transfer. Therefore, the usual time-of-flight observable $n(k)=\langle\hat{n}(k)\rangle$ is flat for both states and one has to analyze higher-order correlations to probe the presence of these states. We are interested in the following quantity :  $\sigma_{\Psi}({\bf k} , {\bf k}') = \langle\hat{n}({\bf k})\hat{n}({\bf k}')\rangle - \langle\hat{n}({\bf k})\rangle\langle\hat{n}({\bf k}')\rangle$, with
\begin{equation}
	\hat{n}({\bf k}) = \frac{|w({\bf k})|^{2}}{N}\sum_{i , j}e^{i{\bf k}.({\bf r}_{i} - {\bf r}_{j})} b^{\dag}_{i}b_{j}\  \  \  \ , \ \ \ w({\bf k}) = e^{-\frac{x^{2}|{\bf k}|^{2}}{4}} \  \
\end{equation}
$x$ defining a phenomenological broadening. The vectors  ${\bf r}_{i}$ defines the position of the sites representing the three wells. These sites are represented as the three summits of an equilateral triangle of length $a=1$ 
. We choose the origin of the coordinates as the center of this triangle.

\subsection{With static repulsive interaction only.}
\begin{figure}[h]
	\centering
	\includegraphics[scale=0.3]{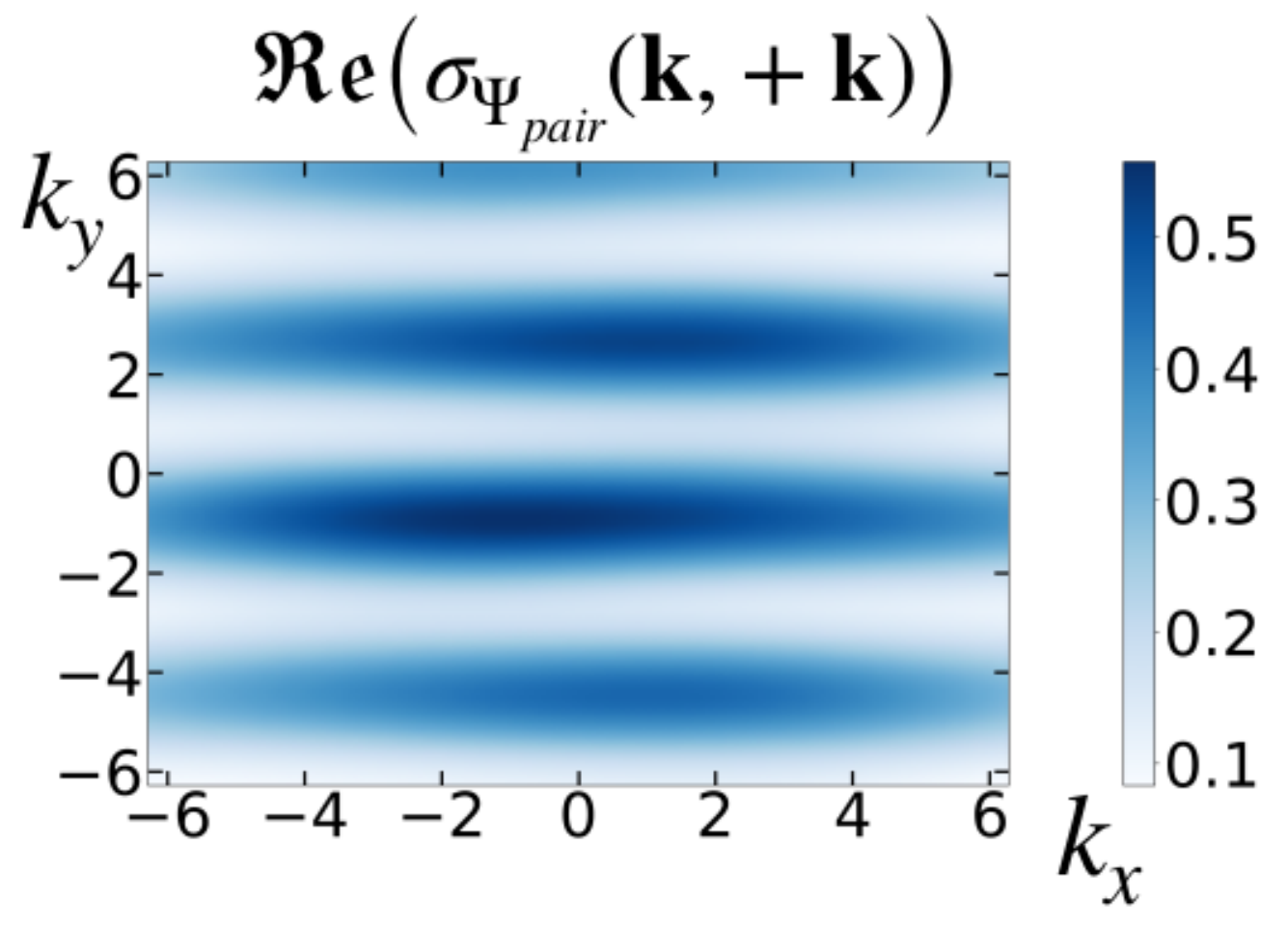}
	\caption{ \small{Characterization of a resonant pair state with a static repulsive interaction. Parameters are $\mu=20, \mu_2=5.781, U_0=0.9$.}}
	\label{fig:map_undriven}
\end{figure}

(See Appendix A) Let us now consider $|\Psi_{pair}\rangle = \alpha|0,N,0\rangle+ \beta|1 ,N-2,1\rangle$.
The two states are connected by operators like $b^{\dag}_{1}b_{2}b^{\dag}_{3}b_{2}$ or $b^{\dag}_{2}b_{1}b^{\dag}_{2}b_{3}$  so that in the case {\bf k}’ = +{\bf k}, the phase acquired is $e^{i{\bf k}.({\bf x}_{21} + {\bf x}_{23})} = e^{i\sqrt{3}k_{y}} $. Finally:
\begin{eqnarray}
	\sigma_{\Psi_{pair}}({\bf k} , +{\bf k}) &=& |w({\bf k})|^{4}\\
\nonumber
\Big[\frac{2}{N} &+& \frac{4\sqrt{N(N-1)}}{N^{2}}|\alpha||\beta|\cos{(\sqrt{3}k_{y} + \phi_{\alpha} - \phi_{\beta})}\Big]
\end{eqnarray}
where $\phi_{\alpha}, \phi_{\beta}$ are the phases of coefficients $\alpha,\beta$.\\
We can see that the resulting pattern is made of stripes modulated along $k_{y}$ with a periodicity $\frac{2\pi}{\sqrt{3}}$ (see Fig. \ref{fig:map_undriven}). Moreover, as $ |0,N,0\rangle $ and $|1 ,N-2,1\rangle$ are connected by only one pair transfer, $\sigma_{\Psi_{pair}}({\bf k} , +{\bf k})$ also depends on the phase difference between the coefficient $\alpha$ and $\beta$. Therefore one can directly probe the coherence of this state by measuring $\sigma_{\Psi_{pair}}({\bf k} , +{\bf k})$.

\subsection{With a driven interaction: ideal cases.}
In the three-mode Bose-Hubbard with a strong offset between the sites 1 and 3, we have shown that the modulation of the inter-particle interaction can dynamically trigger W-type superposition states of the form:
\begin{equation}
|\Psi_{W}\rangle = \alpha|N00\rangle + \beta|0N0\rangle + \gamma|00N\rangle
\end{equation}

One easily checks that the usual time-of-flight observable $n({\bf k})=\langle \hat{n}({\bf k}) \rangle$ is flat. The same holds for different components of $|\Psi_{W}\rangle$ individually and for $|\Psi_{W}\rangle$ itself.
For a 'perfect' state $|\Psi_{W}\rangle$, there is only one contribution to $\sigma_{\Psi_{W}}({\bf k} , {\bf k}')$  which leads to a ${\bf k}$-dependence. This contribution is due to virtual processes where a boson hops from a site $i$ to a site $j$ and comes back so that the phase acquired is $({\bf k} - {\bf k}').{\bf x}_{ij}$. Looking at the case ${\bf k}' = -{\bf k}$ we find :

\begin{widetext}
\begin{equation}
	\mathfrak{Re}\big(\sigma_{\Psi_{W}}({\bf k} , -{\bf k})\big) = \frac{|w({\bf k})|^{4}}{N}\Big[(|\alpha|^{2} + |\beta|^{2}).\cos{\big(k_{x} + \sqrt{3}k_{y})}  + (|\alpha|^{2} + |\gamma|^{2} ) .\cos{\big(2k_{x}\big)} + (|\beta|^{2} + |\gamma|^{2}).\cos{\big(k_{x} - \sqrt{3}k_{y}\big)}\Big]
\end{equation}
\end{widetext}

As the states $|N00\rangle$ , $|0N0\rangle$ and $|00N\rangle$ are not connected by second order hopping, we can write :
\begin{equation}
	\sigma_{\Psi_{S}}({\bf k} , {\bf k}') = \sigma_{|N00\rangle}({\bf k} , {\bf k}')  + \sigma_{|0N0\rangle}({\bf k} , {\bf k}')  + \sigma_{|00N\rangle}({\bf k} , {\bf k}')
\end{equation}
with :
\begin{align}
	\mathfrak{Re\Big[}\sigma_{|0N0\rangle}({\bf k} , -{\bf k})\Big] &= \frac{|w({\bf k})|^{4}}{N}\\
& |\beta|^{2}\Big(\cos{\big(k_{x} + \sqrt{3}k_{y})}  + \cos{\big(k_{x} - \sqrt{3}k_{y}\big)}\Big) \\
	\mathfrak{Re\Big[}\sigma_{|N00\rangle}({\bf k} , -{\bf k})\Big] &= \frac{|w({\bf k})|^{4}}{N} |\alpha|^{2}\Big(\cos{\big(k_{x} + \sqrt{3}k_{y})}  + \cos{\big(2k_{x}\big)}\Big)\\
	\mathfrak{Re\Big[}\sigma_{|00N\rangle}({\bf k} , -{\bf k})\Big]  &= \frac{|w({\bf k})|^{4}}{N} |\gamma|^{2}\Big(\cos{\big(k_{x} - \sqrt{3}k_{y})}  + \cos{\big(2k_{x} \big)}\Big)
\end{align}
Therefore the probabilities in state $W$ can be recovered using the following relations :
\begin{align}
	|\alpha|^{2} &= -\frac{N}{2|w(\frac{\pi}{2} , -\frac{\sqrt{2}\pi}{2})|^{4}}\mathfrak{Re}\Big[\sigma({\bf k},-{\bf k}))\Big]_{{\bf k} = (\frac{\pi}{2} , -\frac{\sqrt{2}\pi}{2})} \\
	|\beta|^{2} &= -\frac{N}{2|w(\pi , 0)|^{4} }\mathfrak{Re}\Big[\sigma\big({\bf k}, -{\bf k})\Big]_{{\bf k} = (\pi ,0)}\\
	|\gamma|^{2} &= -\frac{N}{2|w(\frac{\pi}{2} , +\frac{\sqrt{2}\pi}{2})|^{4}}\mathfrak{Re}\Big[\sigma({\bf k},-{\bf k})\Big]_{{\bf k} = (\frac{\pi}{2} , \frac{\sqrt{2}\pi}{2})} \\
	1 &= -\frac{N}{2}\mathfrak{Re}\Big[\sigma({\bf k},-{\bf k})\Big]_{{\bf k} = (0,0)}
\end{align}

\begin{figure}[h!]
	\centering
	\includegraphics[scale = 0.33]{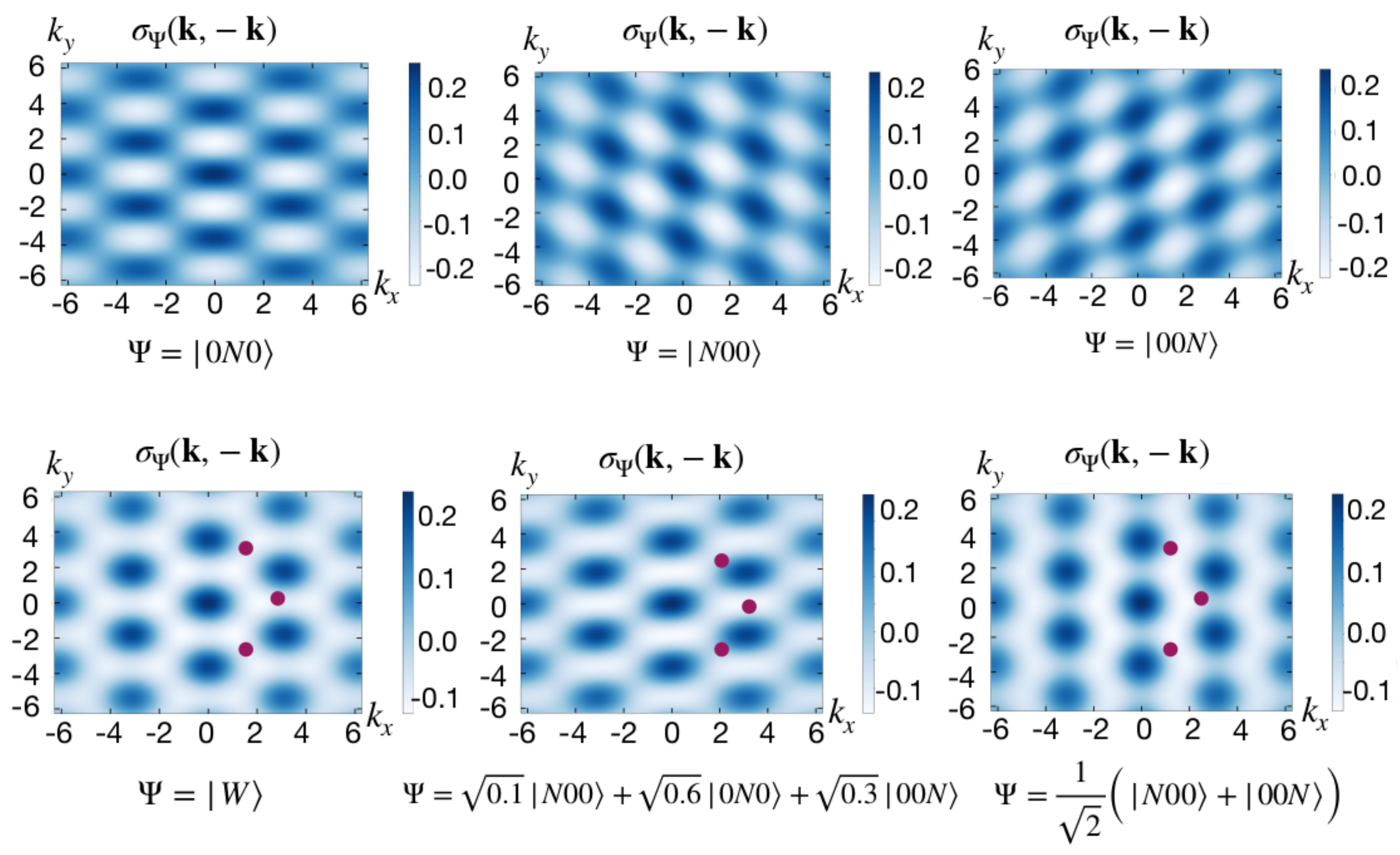}
	\caption{ \small{Readout maps for ideal cases: fully localized (top panels), ideal W- and W-type state, and N00N state (bottom panels). Red dots denote useful points for the determination of the weights of the W-state components (see text).}}
	\label{fig:perfect}
\end{figure}

As we can see in Fig. \ref{fig:perfect}, for a symmetric setup the three Fock states that compose the superposition state give the same pattern but with a different orientation. The perfect W state with $\alpha=\beta=\gamma$ yields a pattern with an hexagonal symmetry and an asymmetric amplitude with respect to zero. A NOON state (for instance $\alpha=\beta, \gamma=0$) and an asymmetric superposition give a pattern similar to that of the perfect W state but distorted in different directions, depending on the coefficients $\alpha$, $\beta$ and $\gamma$. These characteristics allow to probe the presence of a superposition of the three states $|N,0,0\rangle$, $|0,N,0\rangle$ and $|0,0,N\rangle$.

\begin{figure}[ht]
\centering
\includegraphics[scale=0.3]{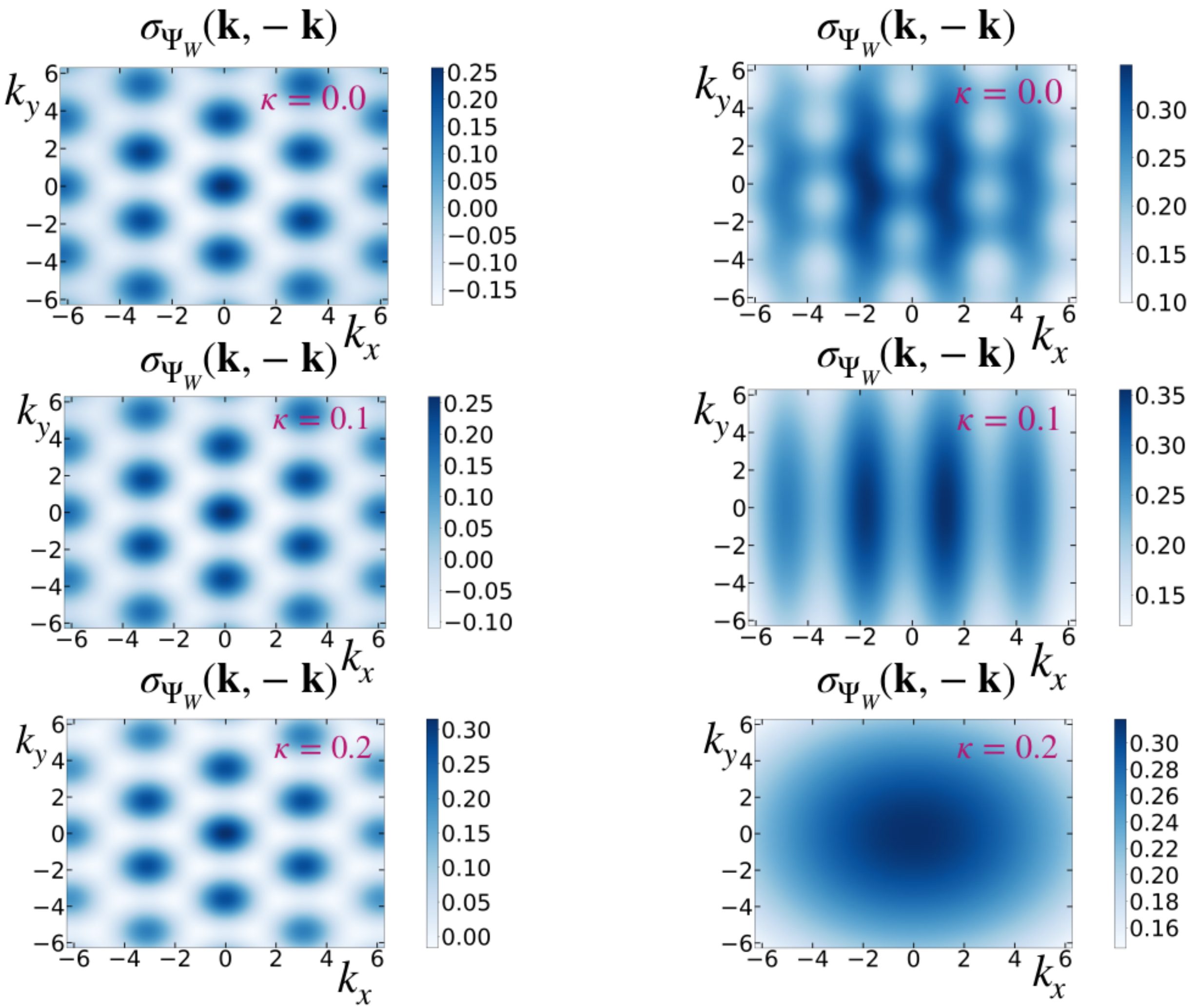}
\caption{ \small{Readout maps for a dynamically generated state, as a function of the filtering parameter $\kappa$ (see text). (Left) $\sigma(k, -k)$ is slightly modified by the parasitic states; (Right) $\sigma(k, k)$ is a direct fingerprint of the parasitic states.}}
\label{fig:parasitic}
\end{figure}

\subsection{Checking purity.}
Here we show some results for N=8 and for superposition states that are generated dynamically : Starting from the state $|0N0\rangle$, we let it evolve and stop at $T = 198 t.J/\hbar$, when the system is in a good superposition state. At this time, the fidelity for such a state is 0.9. This means that the wavefunction has also sizeable components over a few other states.
One finds that  $\sigma_{\Psi_{W}}({\bf k} , {\bf k})$ is flat for a perfect superposition state, this observable can indicate the contributions of the other "parasitic" states. To see this effect, one can delete all the states $ |p\rangle $ in the wavefunction for which the coefficient $|C_p|^{2} < \kappa $ and plot the quantities $\sigma_{\Psi_{W}}({\bf k} , -{\bf k})$ and $\sigma_{\Psi_{W}}({\bf k} , {\bf k})$ for different values of $\kappa$ (Fig. \ref{fig:parasitic}). For  $\kappa = 0.0$, all the states are present, for $\kappa=0.1$ there is a small fraction of the "parasitic states" and for $\kappa = 0.2$ there are only the three states $|N00\rangle$ , $|0N0\rangle$ and $|00N\rangle$.

The parasitic states can also alter the values of $\sigma_{\Psi_{W}}({\bf k} , -{\bf k})$ so that it is more difficult to recover the composition of $|\Psi_{W}\rangle$ with the help of the analytical formula for the perfect state.

In the case  $\kappa= 0.1$, there are only two parasitic states :  $|N-2, 0, 2\rangle$ and $|2, 0, N-2\rangle$. The only contributions to  $\sigma_{\Psi_{W}}({\bf k} , {\bf k})$ are the terms which transfer two particules between 1 and 3, in the same direction, so that the result is an oscillation in the direction $k_x$ (the period is $\pi$, as we can see in Fig. \ref{fig:parasitic}). The two quantities $\sigma_{\Psi_{W}}({\bf k} , {-\bf k})$ and $\sigma_{\Psi_{W}}({\bf k} , {\bf k})$ are thus complementary : the first serves to probe the superposition and the second to check its purity.

As an example, let us consider that $|\Psi\rangle = |\Psi_{W}\rangle + \epsilon \big(|N-2,0,2\rangle + |2,0,N-2\rangle\big)$ like in the case $\kappa = 0.1$.
At order $\epsilon$, the only contribution is obtained when the operator $b^{\dag}_{i}b_{j}b^{\dag}_{l}b_{m}$ connects $|N-2,0,2\rangle + |2,0,N-2\rangle$ with $|\Psi_{W}\rangle$.
There are only two terms so that the result is : $\sigma_{\Psi}({\bf k} , {\bf k}') = \sigma_{\Psi_{W}}({\bf k} , {\bf k}') + 2\epsilon\frac{\sqrt{2N(N-1)}}{N^{2}}\big(e^{i({\bf k} + {\bf k}').{\bf x}_{13}} + e^{i({\bf k} + {\bf k}').{\bf x}_{31}}\big)$.
So in the case ${\bf k} = {\bf k}'$, we can characterize the parasitic states:
\begin{equation}
\nonumber
\sigma_{\Psi}({\bf k} , +{\bf k}) = \sigma_{\Psi_{W}}({\bf k} , +{\bf k}) + 4\epsilon|w({\bf k})|^{4}\frac{\sqrt{2N(N-1)}}{N^{2}}\cos{(2k_{x})} + \mathcal{O}(\epsilon^{2})
\end{equation}

\end{document}